\documentclass{aa}
\usepackage{graphicx}
\usepackage{color}
\usepackage{epstopdf}
\usepackage{txfonts}
\begin{document}
\title{The detached dust and gas shells around the carbon star U~Ant}

\titlerunning{The detached shells around U~Ant}

   \author{M. Maercker  \inst{1},
           H. Olofsson\inst{1,2},
           K. Eriksson\inst{3},
           B. Gustafsson\inst{3},
           F. L. Sch\"oier \inst{2}
	  }
\authorrunning{M. Maercker et al.}

\offprints{M. Maercker}

   \institute{
   	Department of Astronomy, Stockholm University, AlbaNova University Center, SE--106 91 Stockholm, Sweden\\
       	\email{maercker@astro.su.se}
    	\and
     	Onsala Space Observatory,  SE--439 92 Onsala, Sweden
	\and
	Department of Physics and Astronomy, Uppsala University, SE--75120 Uppsala, Sweden 
         }

   \date{Received 30 September 2009 ; accepted December 2009}

 \abstract
{Geometrically thin, detached shells of gas have been found around a handful of carbon stars. The current knowledge on these shells is mostly based on CO radio line data. However, imaging in scattered stellar light adds important new information as well as allows studies of the dust shells.}
{Previous observations of scattered stellar light in the circumstellar medium around the carbon star U~Ant were taken through filters centred on the resonance lines of K and Na. These observations could not separate the scattering by dust and atoms. The aim of this paper is to remedy this situation.}
{We have obtained polarization data on stellar light scattered in the circumstellar medium around U~Ant through filters which contain no strong lines, making it possible to differentiate between the two scattering agents. Kinematic, as well as spatial, information on the gas shells were obtained through high-resolution echelle spectrograph observations of the KI and NaD lines.}
{We confirm the existence of two detached shells around U~Ant. The inner shell (at a radius of $\approx$\,43{\arcsec} and a width of $\approx$\,2{\arcsec}) consists mainly of gas, while the outer shell (at a radius of $\approx$\,50{\arcsec} and a width of $\approx$\,7{\arcsec}) appears to consist exclusively of dust. Both shells appear to have an over-all spherical geometry. The gas shell mass is estimated to be $2\times10^{-3}\,M_{\odot}$, while the mass of the dust shell is estimated to be $5\times10^{-5}\,M_{\odot}$. The derived expansion velocity, from the KI and NaD lines, of the gas shell, 19.5 $\rm{km\,s^{-1}}$, agrees with that obtained from CO radio line data. The inferred shell age is 2700 years. There is structure, e.g. in the form of arcs, inside the gas shell, but it is not clear whether these are due to additional shells.}
{Our results support the hypothesis that the observed geometrically thin, detached shells around carbon stars are the results of brief periods of intense mass loss, probably associated with thermal pulses, and subsequent wind-wind interactions. The separation into a gas and a dust shell, with different widths, is most likely the effect of different dynamical evolutions of the two media after their ejection.}


   \maketitle

\section{Introduction}
\label{intro}

In their final stages of evolution, stars between $\approx0.8-8\,M_{\odot}$ ascend the asymptotic giant branch (AGB). The evolution in this phase is strongly affected by intense mass loss from the stellar surface, with winds corresponding to mass-loss rates in the range $10^{-8}\,M_{\odot}\,\rm{yr^{-1}}$  to $10^{-4}\,M_{\odot}\,\rm{yr^{-1}}$ (Bl\"ocker~\cite{blocker1995}). Although the existence of the mass loss is well established, much remains in the understanding of the mechanism(s) behind it. Present observational uncertainties in the determined mass-loss rates are in the best case a factor of about three (Ramstedt et al.~\cite{ramstedtetal2008}), based on smooth, spherical symmetric wind models. This is unfortunate, as even a moderate change in the mass-loss rate will have a profound effect on the evolution of the star, its nucleosynthesis, and its return of chemically enriched material to the interstellar medium (Forestini \& Charbonnel~\cite{forestinico1997}; Schr\"oder \& Sedlmayr~\cite{schroderco2001}). In particular, the dependances of the mass loss on stellar parameters such as mass, luminosity, radius, temperature, and pulsational properties are essentially unknown.

In general, the mass loss is thought to increase (on average) as the star evolves along the AGB (Habing~\cite{habing1996}).  In addition, temporal variations, on short as well as long time scales, are present and they may be substantial. In a major survey of circumstellar CO radio line emission from nearby carbon stars, Olofsson et al. (\cite{olofssonetal1988}) found two stars with distinctly double-peaked line shapes, suggesting from the star detached gas shells, i.e., an effect of episodic mass loss. The survey was extended, and a few additional objects were detected with similar signs of highly time-variable mass loss [summarized in Olofsson et al.~(\cite{olofssonetal1993}, \cite{olofssonetal1996})]. Maps of the CO($J$\,=\,$1-0$, $2-1$, and $3-2$) emission lines revealed geometrically thin, CO line-emitting shells of gas around the carbon stars R~Scl, U~Ant, S~Sct, V644~Sco, and TT~Cyg (Olofsson et al.~\cite{olofssonetal1996}). High resolution maps, made with the IRAM Plateau de Bure interferometer, of TT~Cyg (Olofsson et al.~\cite{olofssonetal2000}) and the carbon star U~Cam (Lindqvist et al.~\cite{lindqvistetal1999}) showed that the shells are thin ($\Delta R / R < 0.1$), and remarkably spherical. With the detection of a CO line-emitting shell around the carbon star DR~Ser, a total of seven carbon stars with geometrically thin, detached gas shells are known (Sch\"oier et al.~\cite{schoieretal2005}). Detached shells of dust have been observed around  a number of AGB and post-AGB stars (Waters et al.~\cite{watersetal1994}; Izumiura et al.~\cite{izumiuraetal1996},~\cite{izumiuraetal1997}; Speck et al.~\cite{specketal2000}; Wareing et al.~\cite{wareingetal2006}). These dust shells are not necessarily geometrically thin. R~Hya is the only M-type AGB star with a detected detached shell, in this case through dust emission (Hashimoto et al.~\cite{hashimotoetal1998}). The 21-cm HI line emission can be used to study mass-loss rate variations on a longer time scale than that probed by the CO line emission as well as effects of interaction with the ISM (e.g. G\'erard \& Le Bertre~\cite{gerardco2003}). Also in this case shells are found, but they are of a different nature than the detached CO shells.
 
A phenomenon which may affect the mass-loss properties during AGB evolution is the He-shell flash or thermal pulse. During a thermal pulse the star undergoes a structural change induced by explosive He-burning in the He-shell. The thermal pulse causes the surface luminosity, radius, and effective temperature of the star to change (Pols et al.~\cite{polsetal2001}). The most significant features of the pulse are a rapid dip of the luminosity during the pulse, a very brief luminosity peak immediately after the pulse, followed by a slow dip and a subsequent long-lasting period when the star slowly recovers to its pre-pulse luminosity (Wagenhuber \& Groenewegen~\cite{wagenhuberco1998}). The amplitude and duration of the luminosity changes depend mainly on stellar mass and metallicity (Boothroyd \& Sackmann~\cite{boothroydco1988a}). The changes in the stellar structure mix regions with nuclear-processed material in the interior of the star with the surface through convection. This dredge-up leads to a chemical evolution of the stellar envelope, and may result in the formation of a carbon star, changing the C/O-ratio from $<$\,1 (M-type AGB stars) to $>$\,1 (carbon stars). Hence, an understanding of the thermal pulse cycle is essential for understanding AGB evolution. 

Olofsson et al. (\cite{olofssonetal1990}) suggested that the geometrically thin, detached shells seen in CO line emission may be connected to an increase in mass-loss rate caused by the luminosity and temperature changes during a thermal pulse, a scenario developed in more detail by Schr\"oder et al. (\cite{schroderetal1999}). However, Steffen et al. (\cite{steffenetal1998}) showed, using numerical models of the expanding matter, that an increase of mass-loss rate during the thermal pulse alone is not sufficient to form the observed detached gas shells. Instead, the interaction of a faster wind, associated with a brief increase in mass-loss rate, colliding with a previous slower wind  can form geometrically thin shells of gas (Steffen et al.~\cite{steffenetal1998}; Steffen \& Sch\"onberner~\cite{steffenco2000}). This is confirmed by detailed hydrodynamic models of the thermal pulse, the dynamical stellar atmosphere, and the evolution of the expanding medium (Mattsson et al.~\cite{mattssonetal2007}). A variation in mass-loss rate together with a change in the expansion velocity is required to form the observed detached shells. Sch\"oier et al. (\cite{schoieretal2005}) modelled the thermal dust and CO line emission for all known carbon stars with detached shells, and found an increase in shell mass and decrease in shell expansion velocity with increasing shell radius, i.e., increasing shell age, indicating a two-wind interaction scenario where a faster wind sweeps up material from a previous slower wind. Interaction of the stellar wind with the surrounding matter (Libert et al.~\cite{libertetal2007}) or bow shocks between the interstellar medium and the expanding circumstellar envelope (Wareing et al.~\cite{wareingetal2006}) may also lead to the formation of detached shells around some sources, although they are not geometrically thin.

Although powerful in detecting the detached shells of gas around carbon stars, observations of CO radio line emission have their limitations. The emission depends on the excitation and chemistry of the CO molecules, and the translation from brightness distribution to density distribution is not straightforward. Furthermore, the lack of interferometers in the southern hemisphere capable of observing the CO rotational lines limits the possibility to obtain detailed interferometer maps to the few northern sources. Infrared emission from dust provides a probe of the shells, however, these observations are usually of low spatial resolution.

Gonz\'alez Delgado et al.~(\cite{delgadoetal2001}; hereafter GD2001) used a novel technique to study the circumstellar medium around the carbon stars R~Scl and U~Ant. They imaged the circumstellar scattered stellar light through two narrow band (5\,nm) filters (centred on the resonance lines of K and Na). By placing a choronograph over the star, the direct stellar light was avoided and the detached shells could be detected. This allowed for the first time to study the detailed structure of the shells taking advantage of the high angular resolution obtained in optical data. The same technique was used again two years later, this time observing the scattered stellar light from R~Scl and U~Ant in polarisation mode (Gonz\'alez Delgado et al.~\cite{delgadoetal2003}; hereafter GD2003). The major fraction of the polarised light is due to scattering by dust particles, and these observations gave high spatial resolution images of the dust distribution in the circumstellar medium. In the case of U~Ant they proposed a structure of four shells. The inner two shells were tentatively seen only in the images through the filter containing the NaD lines. For the outer two shells they found a difference in polarisation degree. Models of the polarised light indicated that the light from outermost shell is entirely due to dust scattering. However, the bulk of the total scattered intensity ($\approx$70\%) must come from another scattering agent, most likely due to atomic scattering (in the KI and NaD lines).

The observations by GD2001 and GD2003 had the limitation that they could not conclusively separate the different contributions by dust and line scattering. In this paper we present new observations of the circumstellar medium around U~Ant taken with the EFOSC2 (ESO Faint Object Spectrograph and Camera) instrument on the ESO 3.6 m telescope through filters that contain no strong lines, hence making it possible to differentiate between dust- and line-scattered light (new images in the filters centred on the resonance lines of K and Na were also obtained). In addition, we observed U~Ant using the echelle spectrograph EMMI (ESO Multi-Mode Instrument) on the ESO NTT, and obtained a map of the circumstellar medium in CO($J$\,=\,$3-2$) emission with the APEX (Atacama Pathfinder Experiment) telescope. In Sect.~\ref{observations} we present the basic properties of U~Ant as well as describe the EFOSC2, EMMI, and APEX observations. We describe the reduction process in Sect.~\ref{reduction}, and present our results in Sect.~\ref{results}. Finally, we discuss the properties of the circumstellar medium around U~Ant and its origin in Sect.~\ref{discussion}, where we also present our conclusions.


\section{Observations}
\label{observations}

\subsection{U~Ant}
\label{uant}

U~Ant is an N-type carbon star with irregular variability. It is located at a distance of 260 pc (the Hipparcos distance; Knapp et al.~\cite{knappetal2003}). It has a luminosity of 5800~$L_{\odot}$, based on SED fitting at wavelengths $\lesssim10\,\mu$m (Sch\"oier et al.~\cite{schoieretal2005}). The present-day mass-loss rate, based on CO radio line emission models, is low ($\approx\,2\times10^{-8}\,M_{\odot}\,\rm{yr^{-1}}$; Sch\"oier et al.~\cite{schoieretal2005}). Izumiura et al. (\cite{izumiuraetal1997}) estimate the ZAMS mass of U~Ant to be $3-5\,M_{\odot}$, but this must be regarded as highly uncertain since it is based on the time difference between two ejected dust shells assuming they are due to two consecutive thermal pulses.

\begin{table*}[t]
\caption{Observations of the circumstellar environment of U~Ant. The table gives the instrument, the date of observation, the filter (or frequency) used, the filter width, the pixel size (for APEX the beam FWHM is given), the spectral resolution, and the total integration time. }
\label{obssum}
\centering
\begin{tabular}{l l l  c c c c r}
\hline\hline
Instrument & Date 		& filter & $\lambda_{\rm{cen}}$ 	& $\Delta \lambda_{\rm{FWHM}}$ &pixel 	& $\Delta v$ & $t^a$\\
		&			& 	&[nm]Ê& [nm]		& [$\arcsec$]	& [km s$^{-1}$] & [s]\\
\hline
EFOSC2	& Apr. 2002	& Str-y		& 548.2 & 18.2 & 0.32 		& $-$ & 5000\\
		&			& F59		& 589.4 & 5 	& 0.32		& $-$ & 4600\\
		&			& H$\alpha$	& 657.7 & 6.2 	& 0.32		& $-$ & 3900\\
		& 			& F77		& 769.9 & 5 	& 0.32		& $-$ & 90\\
EMMI	& Mar. 2004	& F59		& 589.4 & 5	& 0.33		& 4.7 & 9000\\
		&			& F77		& 769.9 & 5 	& 0.33		& 4.7 & 9000\\
APEX-2a	& Dec. 2006 & $-$ &  345\,GHz & $-$	&18			& 0.5 & 15500\\
\hline\hline
\end{tabular}
\begin{list}{}{}
\item[$^{\rm{a}}$] The total integration time for each polarisation angle is given for the EFOSC2 observations. For the APEX observations the total integration time for the map is given.
\end{list}
\end{table*}

\subsection{EFOSC2 imaging of polarised light}
\label{efosc}

The circumstellar environment of U~Ant was observed in polarised, scattered light during three nights in April 2002 using the EFOSC2 focal reducer camera on the ESO 3.6\,m telescope. The images were taken through an H$\alpha$ filter (657.7 nm), a Str\"omgren y filter (548.2 nm, hereafter Str-y), and narrow filters (5 nm) centred on the resonance lines of Na (589.4 nm, hereafter F59) and K (769.9 nm, hereafter F77), with a pixel scale of 0$\farcs$32/pixel. The average seeing during the three nights was $\approx\,$1$\farcs$3 (see Table~\ref{obssum} for details on the observations and filters used). Since the stellar light outshines the circumstellar scattered light by a factor of $\,\approx\,$$10^4$, the use of a coronographic mask is necessary. A mask with a radius of $\approx\,$4\arcsec~ was chosen, reducing the direct stellar light enough to allow for long exposures without saturating the CCD, while still making it possible to detect scattered light close to the star.

Polarimetric images were taken using the H$\alpha$, Str-y, and F59 filters, using a rotating half-wave plate and a fixed polariser. Images were taken at polarisation angles of 0$^{\circ}$, 45$^{\circ}$, 90$^{\circ}$, and 135$^{\circ}$. The total integration time was 3900\,s/angle, 4600\,s/angle, and 5000\,s/angle in the H$\alpha$, Str-y, and F59 filters, respectively. Standard stars were observed through all four filters (LTT3218 in the H$\alpha$, F59, and F77 filters, and LTT6248 in the Str-y and F77 filters) in order to obtain the absolute flux calibration. In addition, a template star of similar magnitude and spectral type as U~Ant (HD137709) was observed in all filters, in order to characterise the stellar psf, in particular over the area of the circumstellar envelope. An image in direct imaging mode without polarising filter was taken through the F77 filter. However, due to the limited available observing time, only one such image could be taken, with a total integration time of 90\,s. The resulting image is of rather poor quality, and we instead use the results in the F77 filter by GD2003. Finally, in order to determine the total stellar flux, U~Ant was observed in direct imaging mode without the use of a coronograph in all filters.

\subsection{EMMI echelle spectroscopy}
\label{emmi}

In March 2004 we performed spectroscopic observations of U~Ant using the ESO NTT equipped with the echelle spectrograph EMMI. Instead of using a cross disperser to separate the orders, we used the F59 and F77 filters to select the orders which contained the resonance lines. This allowed the use of a long slit that covered the entire shell. 

The slit was placed offset from the star, avoiding (most of) the stellar emission and making it possible to observe the circumstellar scattered light instead. The expansion of the shell causes a splitting of the line. Light from the front part of the shell (as seen from Earth) is shifted to shorter wavelengths, while light from the rear part is shifted to longer wavelengths. The velocity difference between the two parts is largest where the line of sight goes through the middle of the shell, where the expansion along the line of sight is largest, while it is zero at the top and bottom edges of the shell. Since the circumstellar envelope is dominated by a spherical, geometrically thin, expanding shell, this results in very characteristic elliptical shapes. The length of the observed ellipse gives a measure of the size of the circumstellar shell, while the width of the ellipse measures the expansion velocity of the shell (see Sect.~\ref{expvel}). 

Spectra were taken at slit offsets of 15\arcsec\, east and west, and 25\arcsec\, east of the star. The total integration time was 9000\,s in each filter at each offset. The average seeing during the night was $\approx\,$1\arcsec. The pixel scale of EMMI is 0$\farcs$33/pixel. We used an 0$\farcs$8 wide slit, resulting in a spectral resolution of R$\,\approx\,$63500, or $\approx\,$4.7\,$\rm{km\,s^{-1}}$.

The standard star HR718 was unfortunately only observed during an observing run one year earlier when the same instrumental set-up was used. This, in addition to the non-standard instrumental set-up of the observations, made a reliable flux-calibration impossible.

\subsection{APEX CO radio line emission}
\label{apex}

An on-the-fly (OTF) map of the U~Ant circumstellar envelope in the CO($J$\,=\,$3-2$) line at 345\,GHz was obtained with the APEX telescope equipped with the APEX-2a receiver during December 2006. Due to the limited amount of observing time, we observed only the south-eastern quadrant in a $70\arcsec\times70\arcsec$ map. The total integration time of the map is 4.3 hours, resulting in an rms noise level of $\approx\,$0.1\,K at a resolution of 0.5\,$\rm{km\,s^{-1}}$. The beam width of APEX at 345\,GHz is 18\arcsec.


\section{Data reduction}
\label{reduction}

\subsection{EFOSC2 images}
\label{EFOSCred}

The EFOSC2 images were reduced using standard tasks for CCD reduction in IRAF (in particular the \emph{ccdproc} task in the \emph{imred.ccdred}~package). The images were bias-subtracted, flat-fielded, and average-combined after aligning the individual images using background stars. In order to subtract the stellar psf from the final images, it is important to know the exact position of the central star behind the occulting mask. The position was determined by finding the centre of rings of constant intensity close to, but outside of, the occulted region. The position was confirmed by measuring the distance to background stars in the image. 

The images were flux calibrated using the observed standard stars. The surface brightness of the scattered light was calculated by determining the scale factor $f_{\rm{scale}}$ that converts $[\rm{counts\,s^{-1}}\,pix^{-1}]$ to $[\rm{erg\,s^{-1}\,cm^{-2}\,\arcsec^{-2}}]$. The scale factor was calculated as 

\begin{equation}
\label{EFOSCcali}
f_{\rm{scale}}={{F_{\rm{Std}}\times\Delta \lambda}\over{F_{\rm{cnts}}\times pixscale^2}},
\end{equation}

\noindent
where $F_{\rm{Std}}$ is the tabulated flux density of the standard star in [$\rm{erg\,s^{-1}\,cm^{-2}\,\AA^{-1}}$], $\Delta \lambda$ the filter width in $\AA$, $F_{\rm{cnts}}$ the number of counts per second in the image of the standard star (determined by a gaussian fit to the observations), and $pixscale$ the pixel scale of the image in arcsecs. The flux calibration is estimated to be accurate within a factor of two. The subtraction of the template star, however, may lead to significant uncertainties in the determined fluxes (see Sect.~\ref{polan}). The stellar fluxes were determined by directly comparing the number of counts per second in the standard star with the tabulated flux and using this factor to convert from counts per second to flux in the direct imaging frames of U~Ant.

\subsection{Polarisation with EFOSC2}
\label{polan}

Several different sources may contribute to the measured polarised light: the central star, stellar light scattered in the interstellar medium, the sky, and the telescope and instruments, as well as polarised light from the sky background. In the following analysis, however, we assume that the stellar light is unpolarised. This is confirmed by \emph{BVRI} polarimetry studies of variable stars, setting a strong upper limit on the polarisation of the stars ($\lesssim1\%$, Raveendran~\cite{raveendran1991}). Further, the relative proximity of U~Ant suggests a negligible contribution to the observed polarisation from the interstellar medium. The observations were made just after new moon, limiting the amount of polarised light due to the background. Finally, although time limitations prevented us from taking images of polarimetric standards in order to characterise the effects of the telescope and the instrument, there are no indications that these might introduce any significant polarisation in the observations. This was confirmed by GD2003, who observed template stars using the same instrumental set-up. They found no polarisation in the regions where the scattered stellar light was detected.  

Therefore, assuming that all the polarised light is due to scattering in the circumstellar environment of the star, the Stokes parameters for the shell emission are given by

\begin{equation}
\label{stokesparams}
\left\{
\begin{array}{l}
I_{sh}=I_m-I_{st}-I_{bg}\\
Q_{sh}=Q_m\\
U_{sh}=U_m\\
P_{sh}=(Q_{sh}^2+U_{sh}^2)^{0.5},\\
\end{array} \right.
\end{equation}
\noindent
where $I_m$, $I_{st}$, and $I_{bg}$ are the total measured intensity, the total intensity due to stellar light scattered in the interstellar medium, Earth's atmosphere and the instrument, and the total intensity in the sky background, respectively. $Q_m$ and $U_m$ are the measured Stokes' parameters, and $P_{sh}$ is the total polarised flux in the shell. The intrinsic polarisation degree of the shell is given by

\begin{equation}
\label{poldegree}
p_{sh} = {{P_{sh}} \over {I_{sh}}} = {{(Q_{sh}^2+U_{sh}^2)^{0.5}}Ê\over {I_{sh}}}.
\end{equation}

The Stokes' parameters can now be determined in a straightforward way from the observed images

\begin{equation}
\label{stokesframes}
\left\{
\begin{array}{l}
I_{m}=F(0^{\circ})+F(90^{\circ})\\
Q_{m}=F(0^{\circ})-F(90^{\circ})\\
U_{m}=F(45^{\circ})-F(135^{\circ}),\\
\end{array} \right.
\end{equation}
\noindent
where $F$ indicates the frames taken at the different polarisation angles. 

\begin{figure}[t]
   \centering
   \includegraphics[width=8cm]{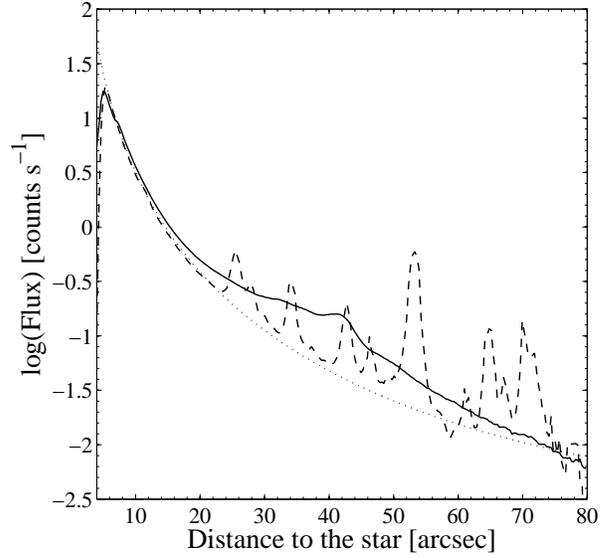}
      \caption{Azimuthally averaged radial profiles (AARPs) in the F59 filter of U~Ant (solid line), the template star (dashed line), and a moffat function fitted to the data (dotted line).}
              \label{Nascale}
    \end{figure}

The most difficult task is the determination of the total intensity in the circumstellar shell $I_{sh}$. The stellar psf, due to light scattered in the Earth's atmosphere and the telescope, was first removed by scaling and subtracting an image of a template star with the same spectral characteristics as U~Ant. This requires an exact alignment of the template and source stars, as well as a very well-defined and `clean' template psf. Unfortunately, the images of the template star contained a large number of back- and foreground stars, making it impossible to cleanly subtract the stellar psf. We therefore fitted Moffat profiles (Moffat~\cite{moffat1969}) to the azimuthally averaged radial profiles (AARPs) of the brightness distributions in the U~Ant template star images (the latter were averaged over the same position angles as in the U~Ant images).  The Moffat profile is described by

\begin{equation}
\label{moffat}
I=I_0\times[1+({{r}\over{r_0}})^2]^{-b},
\end{equation}
\noindent
where $I_0$ is the maximum intensity, $r$ the distance from the centre, and $r_0$ the width of the profile. The exponent $b$ determines how quickly the profile decreases. The psf is assumed to be circularly symmetric. The best-fit Moffat profile is determined manually taking both the U~Ant and the template star profiles into consideration. The exponent, width, and maximum intensity of the Moffat profile were adjusted until it matched the U~Ant observations close to the edge of the occulting mask ($\lesssim10\arcsec$, where the stellar flux dominates over the circumstellar scattered light), and the tail of the psf well outside of the circumstellar region ($\gtrsim70\arcsec$). Figure~\ref{Nascale} shows the AARPs of the total observed intensity in U~Ant and the template star, and the fitted Moffat profile in the F59 filter. The Moffat profile fit to the template star works reasonably well, and the excess emission due to the circumstellar envelope in the U~Ant data can clearly be seen. However, the fit to the U~Ant data is not as good at large distances from the star, and the shape of the AARP of the total shell intensity is very sensitive to the fit of the moffat function.

The uncertainty of the template subtraction and other instrumental effects significantly reduce the quality of the information in the central parts of the images. As reliable measurements are not possible, this region is blanked out in the images by a mask with a 20\arcsec radius. The dominant diffraction spikes due to the spiders were blanked out as well.

\subsection{EMMI echelle spectra}
\label{emmian}

The spectroscopic images from EMMI were reduced using standard tasks in IRAF for flat fielding and bias subtraction. The spectrum obtained on the CCD is usually significantly curved when a long slit is used. This curvature is not necessarily constant over the entire CCD, and the spectrum needs to be carefully straightened. This was done using different tasks for long-slit spectroscopy in the IRAF \emph{noao.twospec.longslit} package.

Due to the use of order-sorting filters instead of a cross disperser, the determination of the dispersion in the EMMI data proved to be problematic. In fact, three lines were visible in the F77 data, although at most two (the two resonance lines) were expected, indicating an order overlap. This was confirmed by examining the profiles of the flat-field images along the dispersion axis, showing that the F77 filter has its centre in the region between two orders. Hence, the lines observed in the F77 filter show two orders of the 769.9 nm line, and one order of the 766.5~nm line. Although the dispersion usually does not vary much between neighbouring orders in echelle spectra, the widths of the ellipses (due to the expansion of the shell) in the two orders differs significantly (by $\approx25\%$). The widths of the 769.9 nm and 766.5 nm lines in the same order differ as well, indicating that the dispersion varies also within one order. The two orders lie close to the edge of the CCD, hence the dispersion may be affected by the optics of the instrumental setup. A careful determination of the dispersion at the position of the lines is therefore extremely important for a correct determination of the kinematical properties. 

The identification of spectral lines in the calibration spectra is nearly impossible when different orders overlap. However, although the images were taken well offset from the central star, stellar light scattered in the atmosphere and the instrument is still projected onto the slit. This, in principle unfortunate fact, made it possible to compare the stellar spectrum with an archived and wavelength-calibrated spectrum of a star of similar spectral type (the carbon star HD20234), resulting in a good dispersion calibration. 

The centre of the F59 filter lies close to the centre of one order, hence there is little or no order overlap in this filter. In this case, however, the calibration spectrum is dominated by one very strong line, making it impossible to identify other lines in the spectrum for a wavelength calibration. [In the case of ordinary echelle spectra, this does not create a problem, as the dispersion solution is extrapolated from the calibration in neighbouring orders.] The atmospheric Na resonance lines are clearly visible in the data though, and the comparison with a spectrum of a star of similar spectral type leads to a good dispersion solution also in this filter.  

\subsection{APEX CO radio line data}
\label{apexred}

\begin{figure}[t]
   \centering
   \includegraphics[width=7cm,angle=-90]{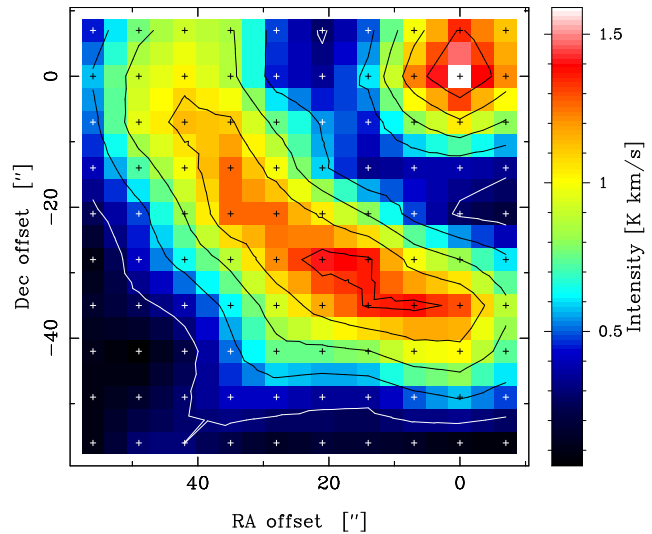}
         \caption{A CO($J$\,=\,$3-2$) velocity-integrated intensity map in a $\pm$\,5\,km\,s$^{-1}$ interval centered on the systemic velocity. U~Ant is located in the (0,0)-position and the map covers the second quadrant of the detached shell. Emission from the detached shell as well as the present-day mass-loss wind is apparent. The contours are based on a grid-spacing of 8\arcsec, while the colour image has an interpolated grid spacing of 4\arcsec. The contour levels are 0.3, 0.6, 0.9, 1.2, and 1.4 K\,km\,s$^{-1}$.}
       \label{uant_map}
    \end{figure}

The on-the-fly data was reduced using {\sc class}\footnote{Continuum and Line Analysis Single-dish Software from the Observatoire de Grenoble and Institut de Radio-Astronomie Millim\'etrique }. The entire data set was inspected manually, and about 80 spectra out of 735 were dropped because of various anomalies. The baselines in the remaining data were relatively stable, and only second order polynomial fits were subtracted. Once we had a consistent data set, the remaining spectra were averaged per position, using a weight determined by the noise level, to a grid-spacing of 8\arcsec. Figure~\ref{uant_map}~shows the map of the velocity-integrated intensity, in a $\pm$\,5\,km\,s$^{-1}$ interval centered on the systemic velocity (this range is chosen since it gives the best view of the shell position, considering also the signal-to-noise ratio), in the second quadrant of the detached shell. The shell can clearly be seen and its location is consistent with the results from the CO($J$\,=\,$1-0$ and $2-1$) maps reported by Olofsson et al. (\cite{olofssonetal1996}). Emission from the present-day mass loss is also visible. The decrease of emission to the north and west is an artefact due to the limited map size in combination with the beam size.

\section{Results}
\label{results}

\subsection{Review of previous results}

\begin{figure}[t]
   \centering
   \includegraphics[width=8cm]{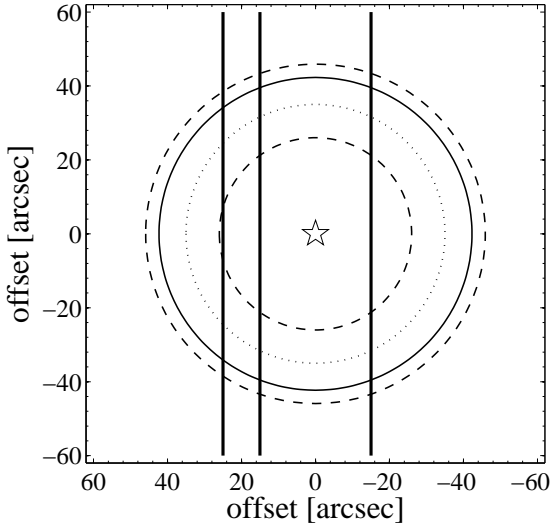}
         \caption{A schematic drawing showing the positions of shells around U~Ant as derived in GD2001 and GD2003 (shells 1 to 4 in order of increasing size; shells 1 and 2 were tentatively introduced). The vertical lines indicate the positions of the slit in the EMMI observations presented here.}
       \label{shells_graph}
    \end{figure}

CO radio line observations of U~Ant show a, not so common, highly double-peaked profile. This was interpreted as emission from a geometrically thin, detached gas shell around U~Ant, an interpretation confirmed by mapping the CO line emission (Olofsson et al.~\cite{olofssonetal1996}). Models show that the CO data is consistent with a detached shell of gas with a radius of $\approx43$\arcsec~(Sch\"oier et al.~\cite{schoieretal2005}). GD2001 and GD2003 observed the circumstellar environment of U~Ant in both direct imaging and polarisation mode through two filters covering the Na and K resonance lines (the F59 and F77 filters, respectively). They introduced four shells, at $\approx$\,25{\arcsec}, 37{\arcsec}, 43{\arcsec}, and 46{\arcsec} (shells 1 to 4, respectively). Shells 1 and 2 were only tentative. Shell 4 was introduced tentatively in GD2001, and it was confirmed in GD2003. The polarisation measurements of GD2003 indicate a separation of the dust and the gas, shell~4 consisting of mainly dust while shell 3 is dominated by gas. The CO radio emission lines most likely come from shell 3. Figure~\ref{shells_graph} shows the positions of the different shells. The vertical lines show the positions of the slit in the EMMI data presented here. Izumiura et al.~(\cite{izumiuraetal1997}) reported the detection of two dust shells in IRAS images at 60$\,\mu$ and 100$\,\mu$m. They processed the images to obtain higher-resolution IRAS (HIRAS) images and derived shell radii of $\approx46\arcsec$ and $\approx3\arcmin$, respectively. The shell at 3{\arcmin} is outside our field of view, and will not be considered further.

In light of these previous results, we present the results of our new data in the following sections, describing the images in polarised, scattered light from EFOSC2 (Sect.~\ref{pollight}) and the derivation of shell fluxes, radii and widths (Sect.~\ref{shellstruct}). The expansion velocity of shell 3 is determined from the EMMI data (Sect.~\ref{expvel}). The different contributions to the scattered light from the dust and gas are estimated (Sect.~\ref{dustgas}) and the new CO data from APEX is modelled (Sect.~\ref{shell3CO}). Finally, the dust masses in the individual shells are determined (Sect.~\ref{masses}).

\subsection{Images and radial profiles of the scattered light}
\label{pollight}

\begin{figure*}[t]
   \centering
   \includegraphics[width=6cm]{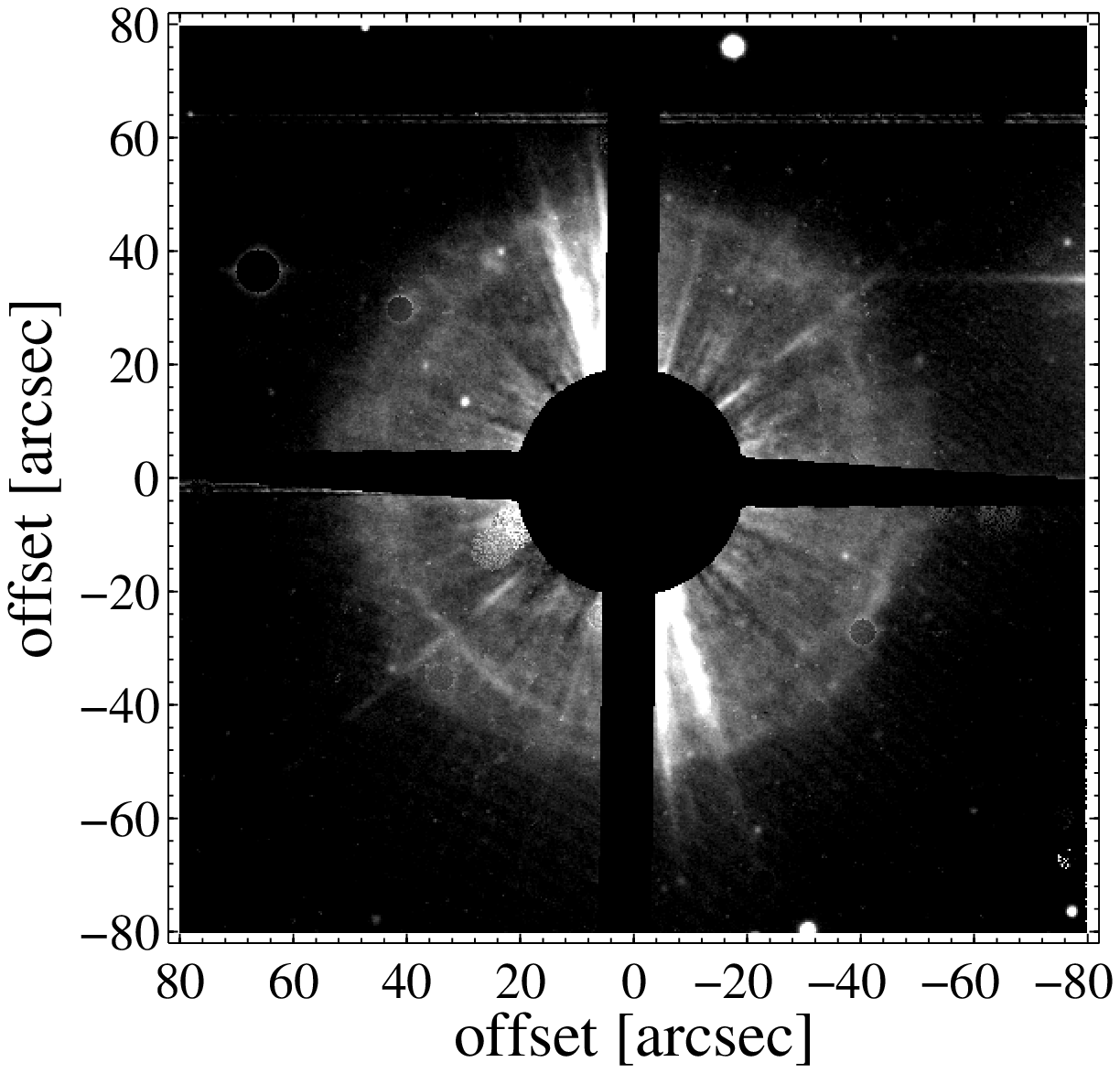}
   \includegraphics[width=6cm]{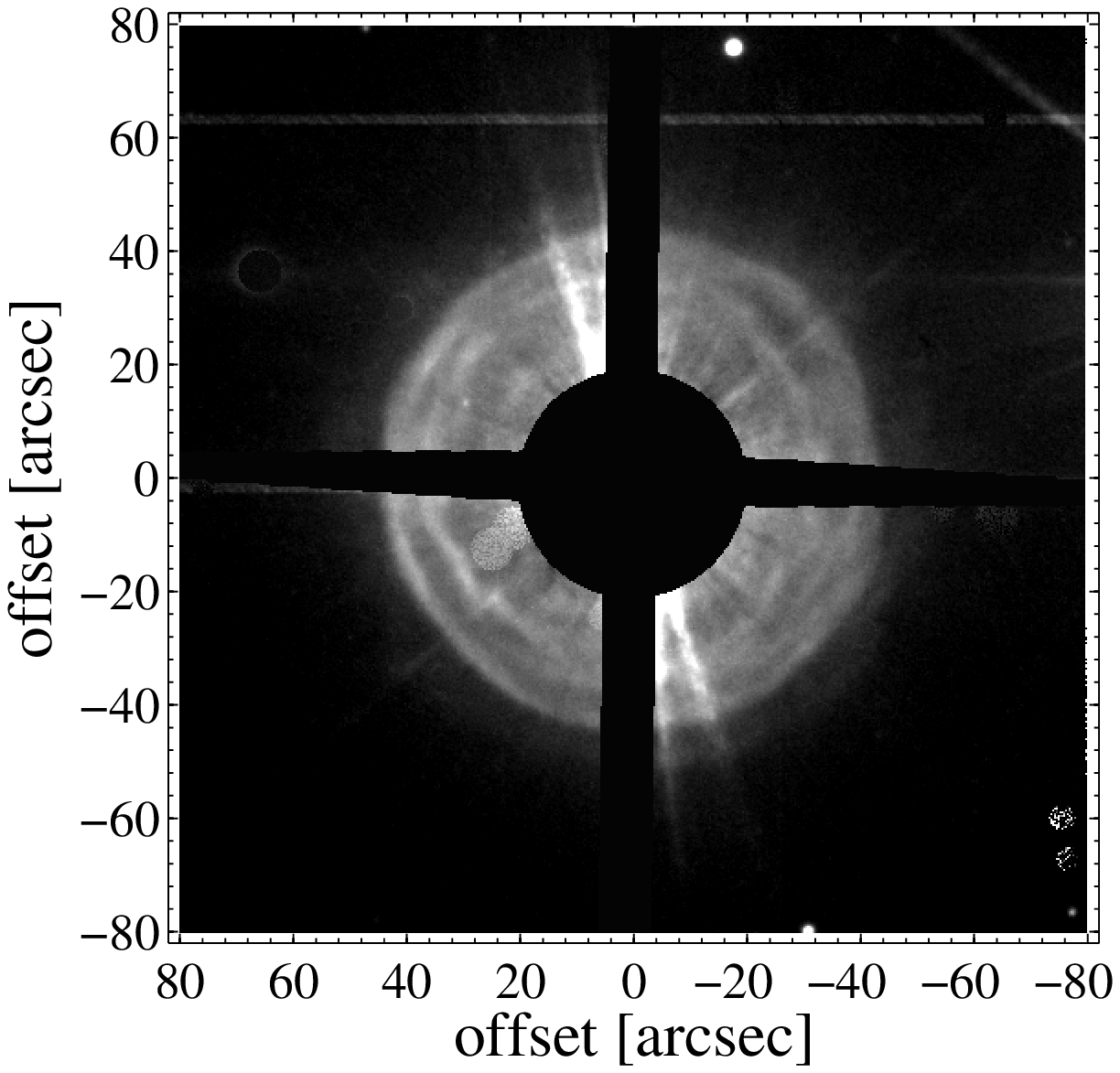}
   \includegraphics[width=6cm]{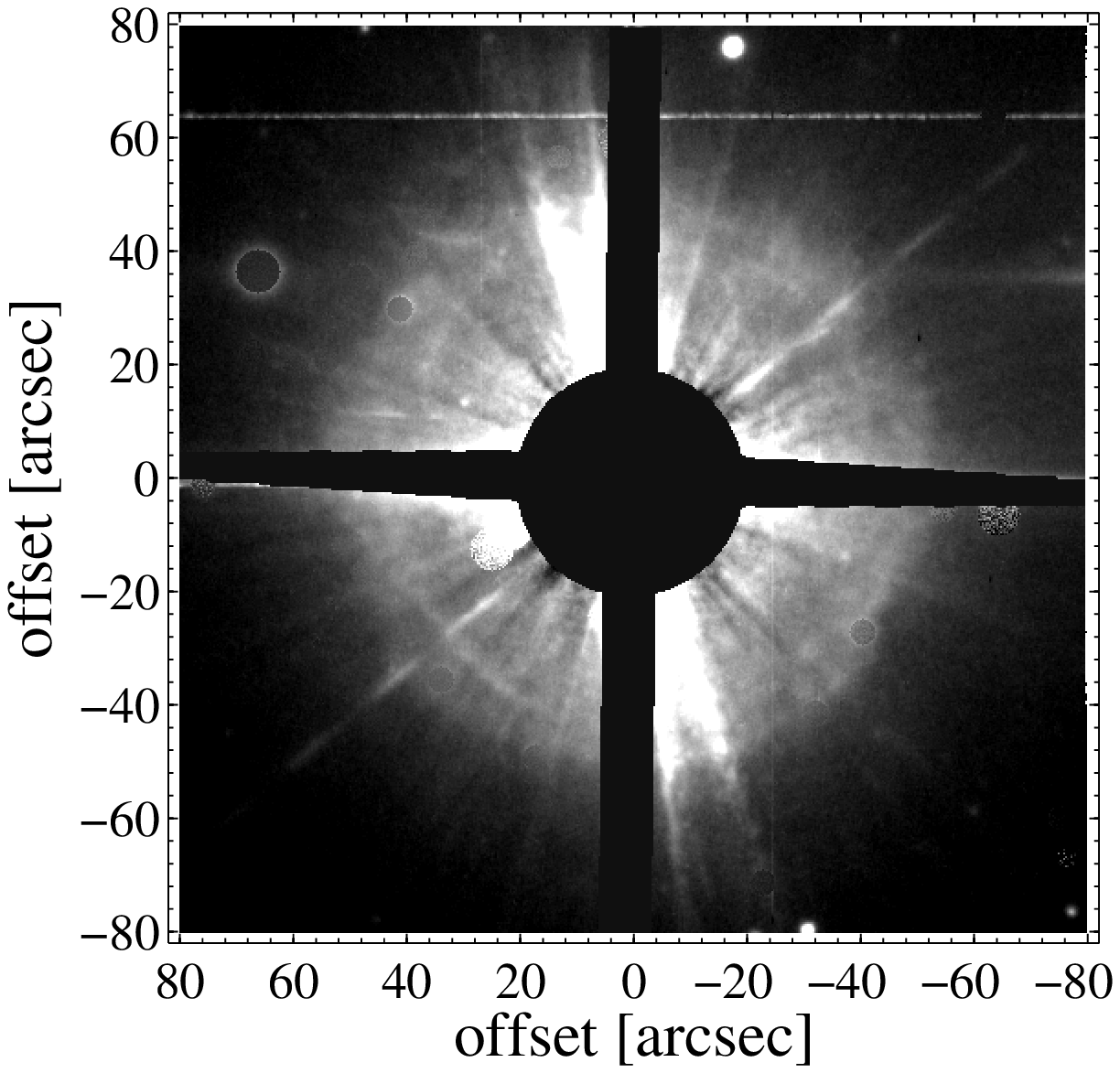}
   \includegraphics[width=6cm]{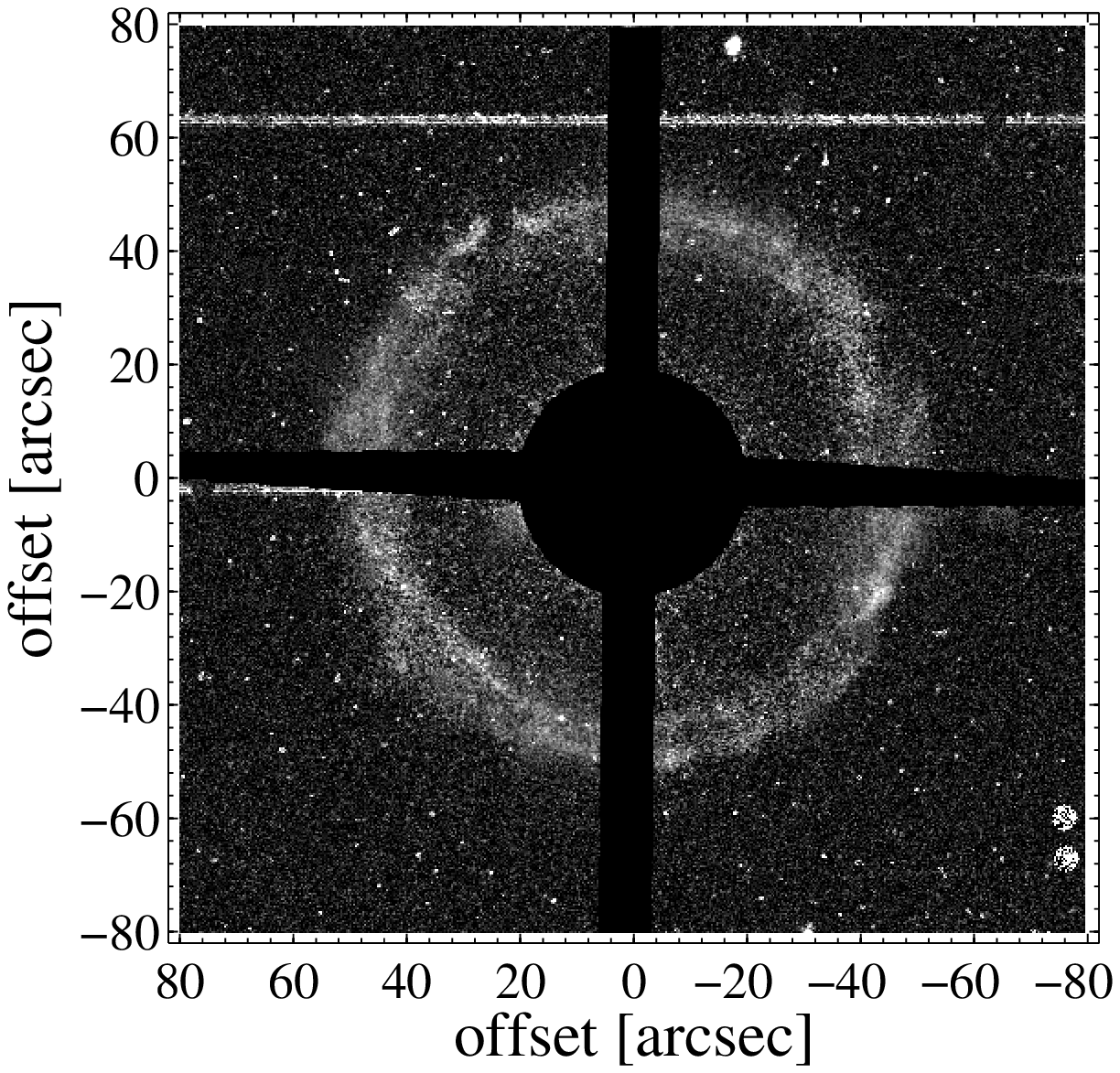}
   \includegraphics[width=6cm]{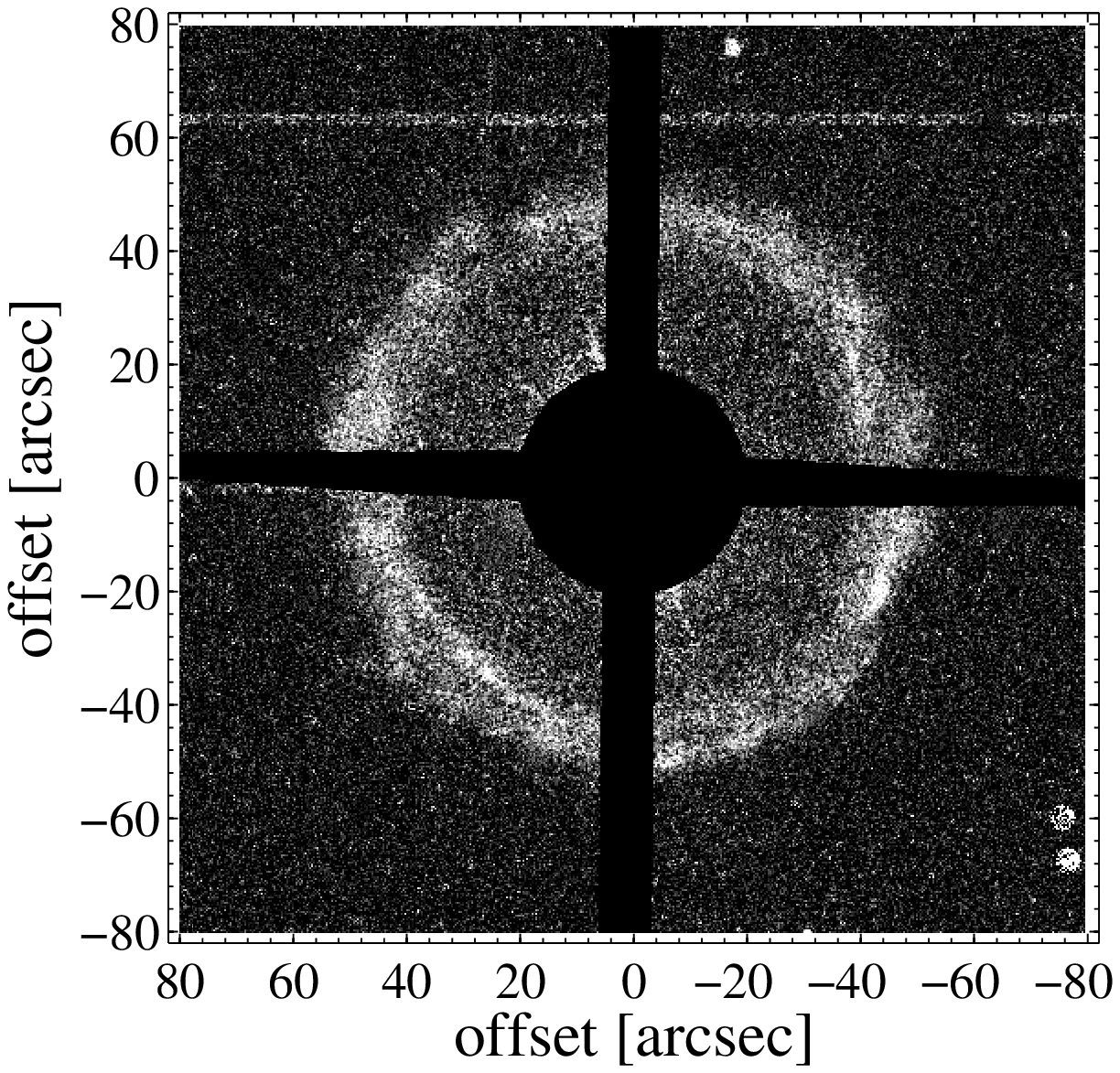}
   \includegraphics[width=6cm]{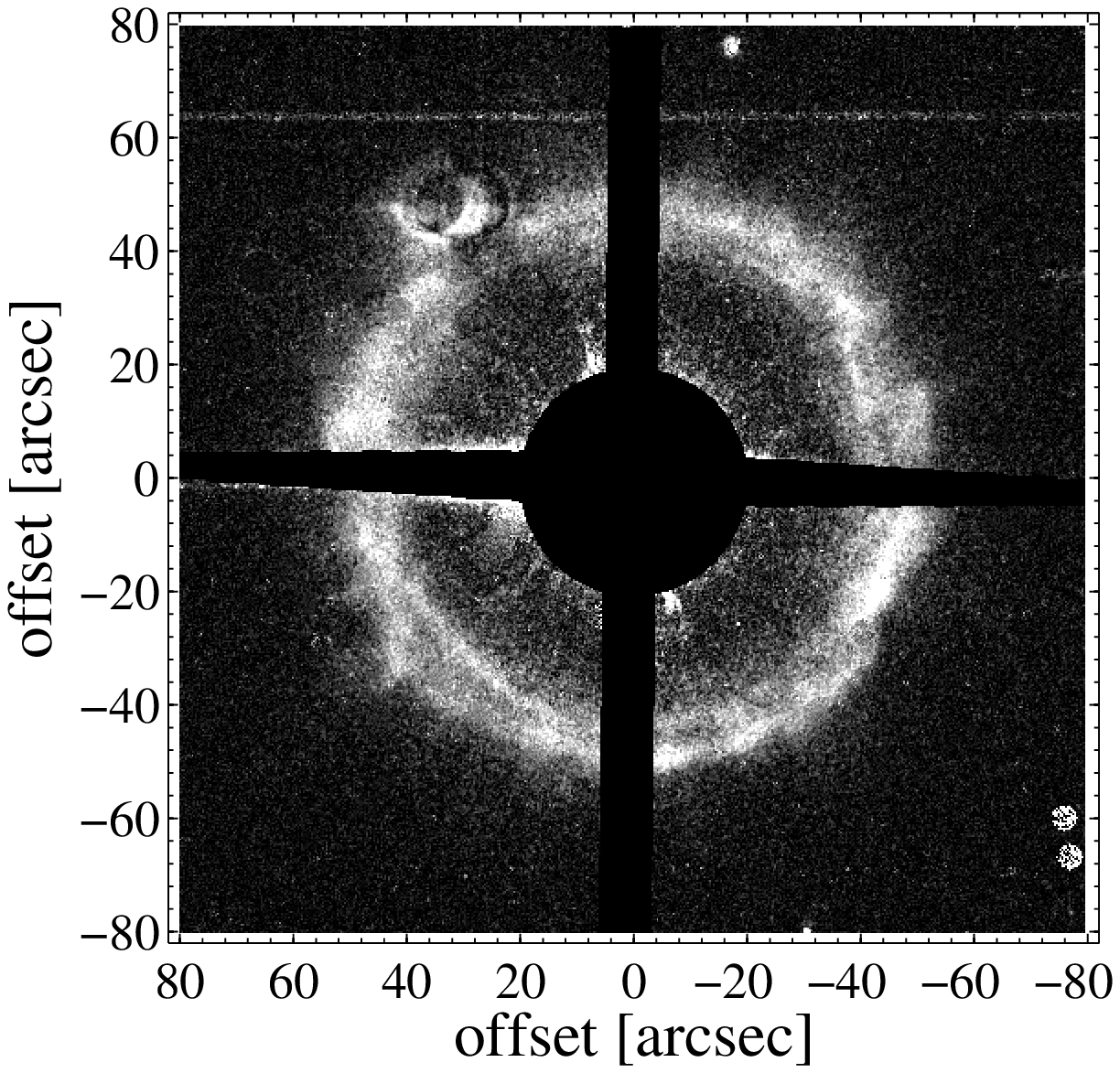}
      \caption{The results of the EFOSC2 observations of U~Ant. \emph{Top row, left to right:} the total intensity ($I_{sh}$) in the Str-y, F59, and H$\alpha$ filters, respectively. \emph{Bottom row, left to right:} the polarised intensity ($P_{sh}$) in the Str-y, F59, and H$\alpha$ filters, respectively.}
       \label{EFOSC_images}
    \end{figure*}

\begin{figure*}[t]
   \centering
   \includegraphics[width=6cm]{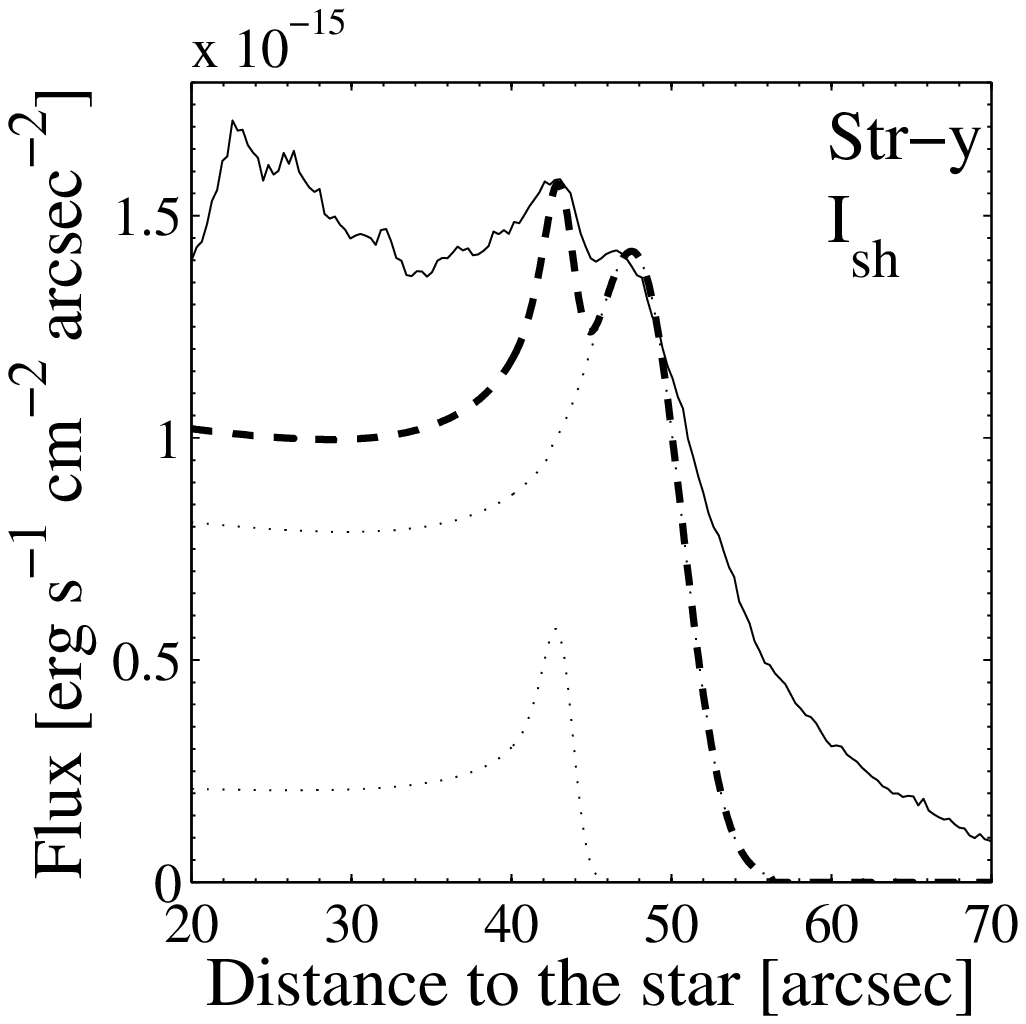}
   \includegraphics[width=6cm]{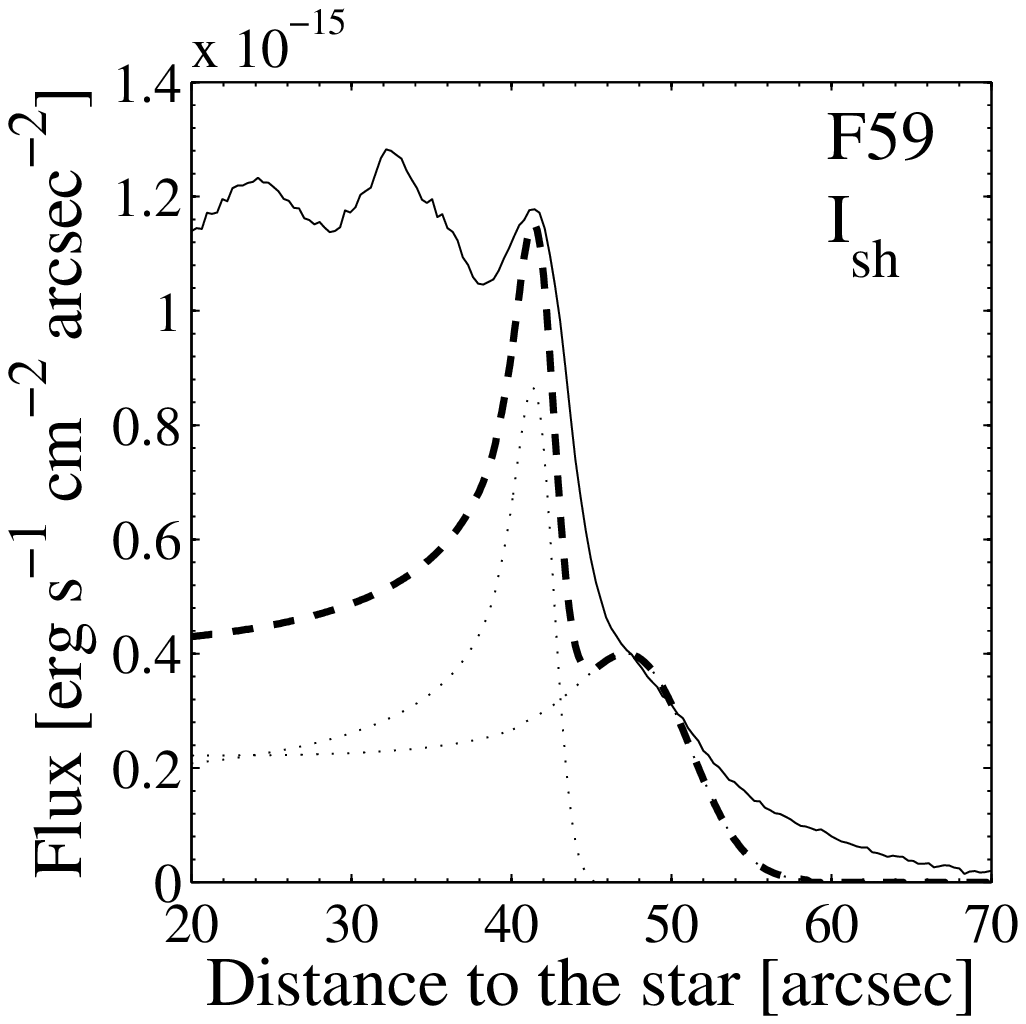}
   \includegraphics[width=6cm]{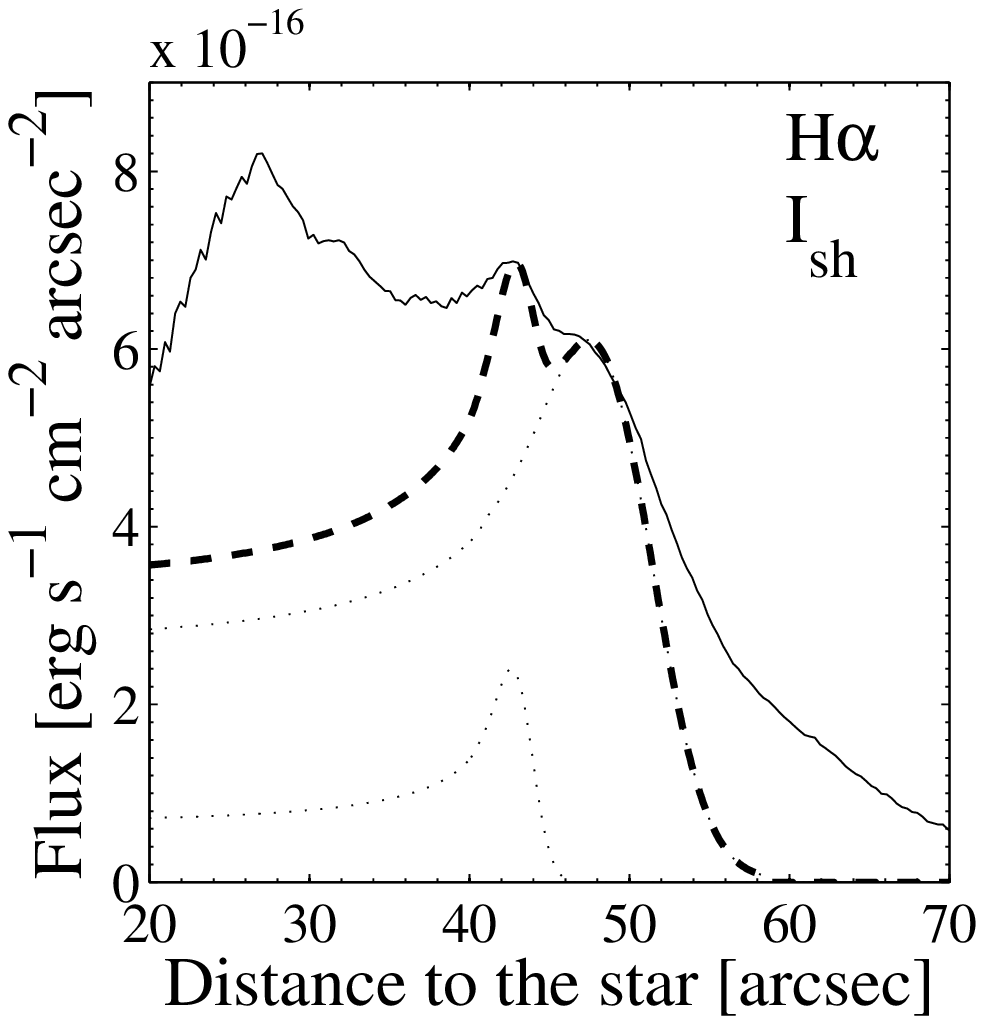}
   \includegraphics[width=6cm]{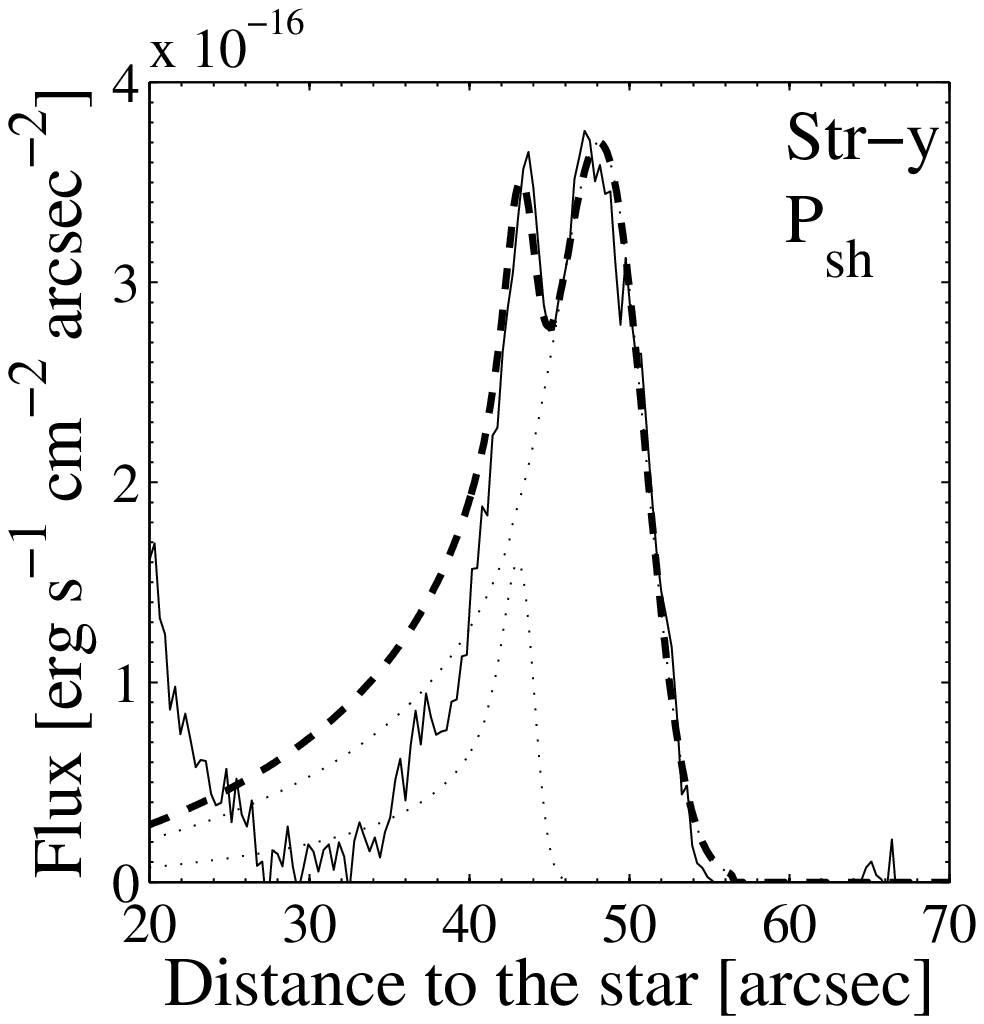}
   \includegraphics[width=6cm]{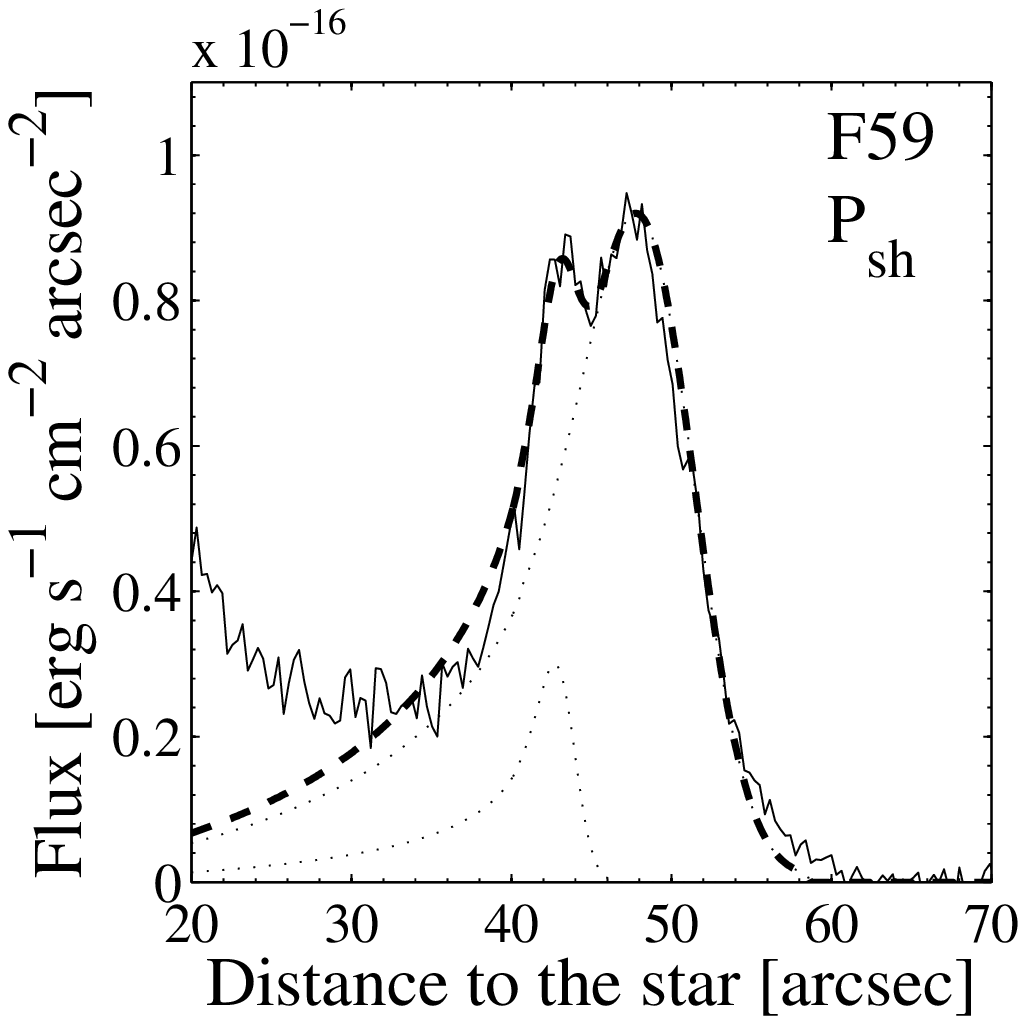}
   \includegraphics[width=6cm]{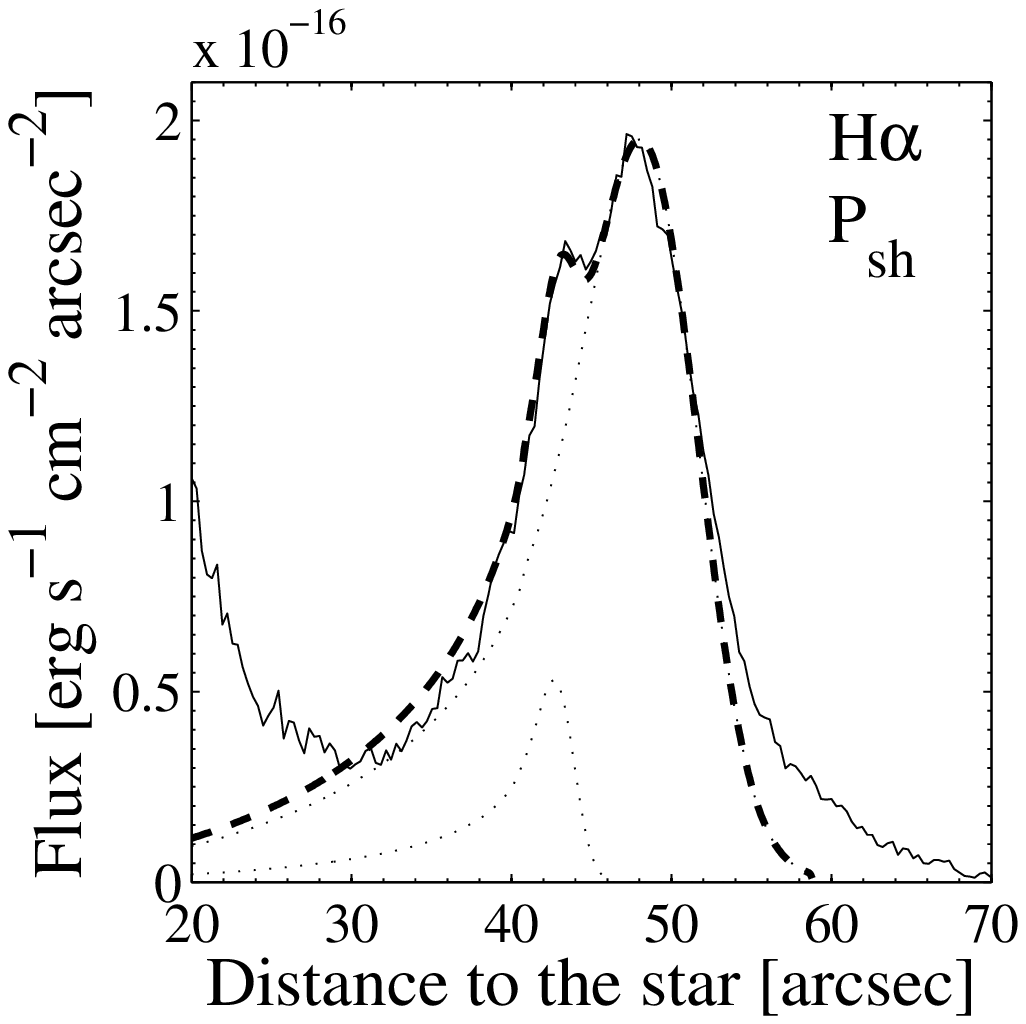}
   \includegraphics[width=6cm]{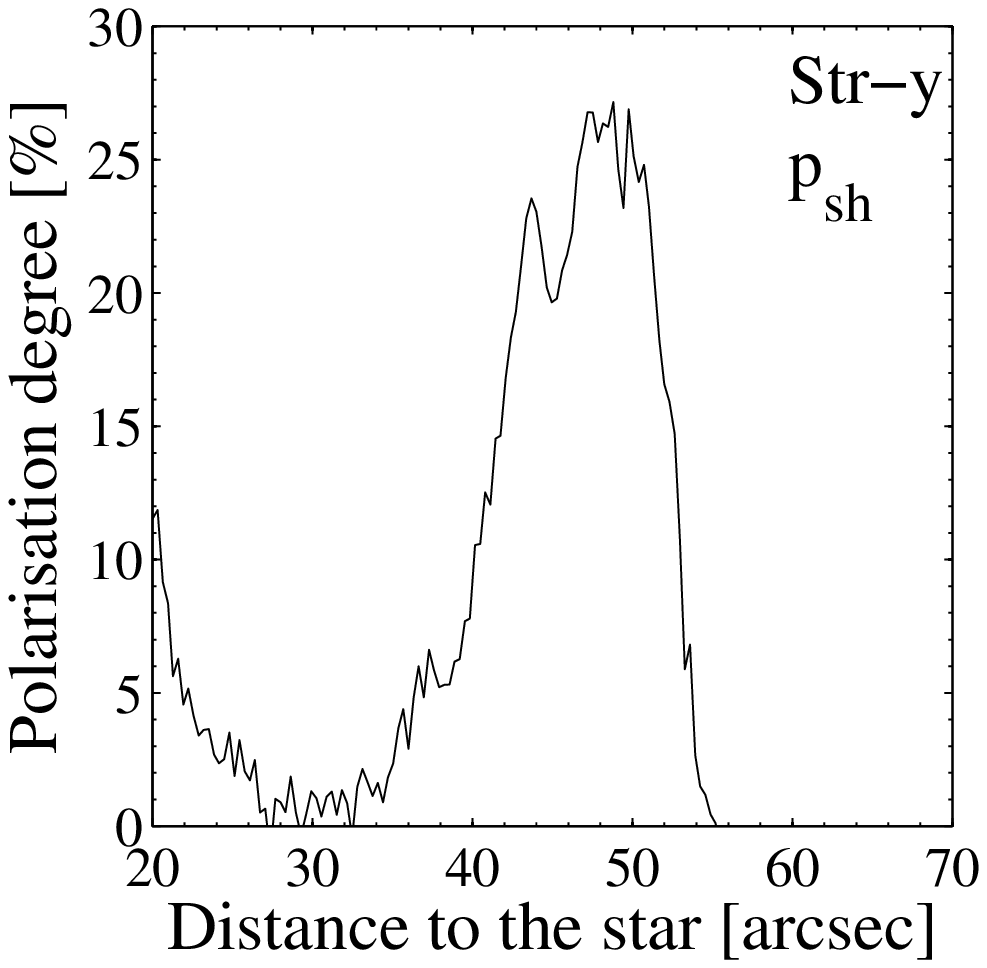}
   \includegraphics[width=6cm]{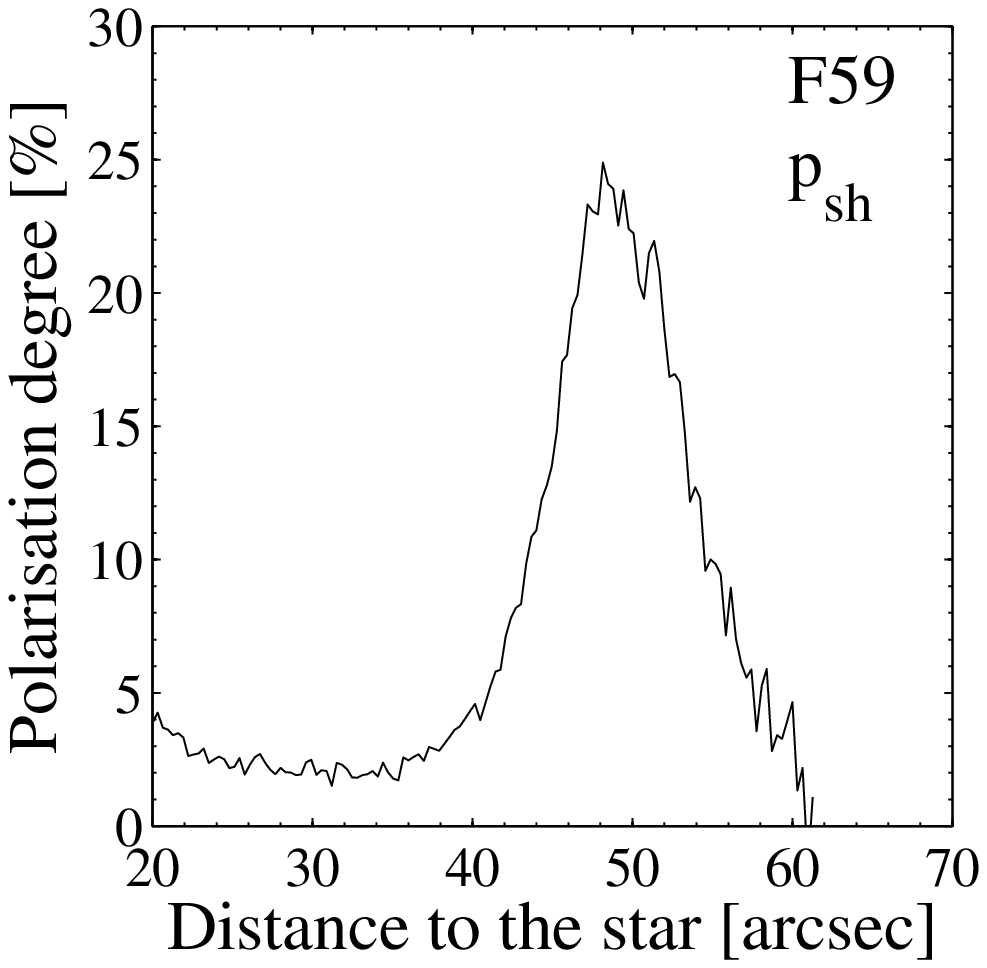}
   \includegraphics[width=6cm]{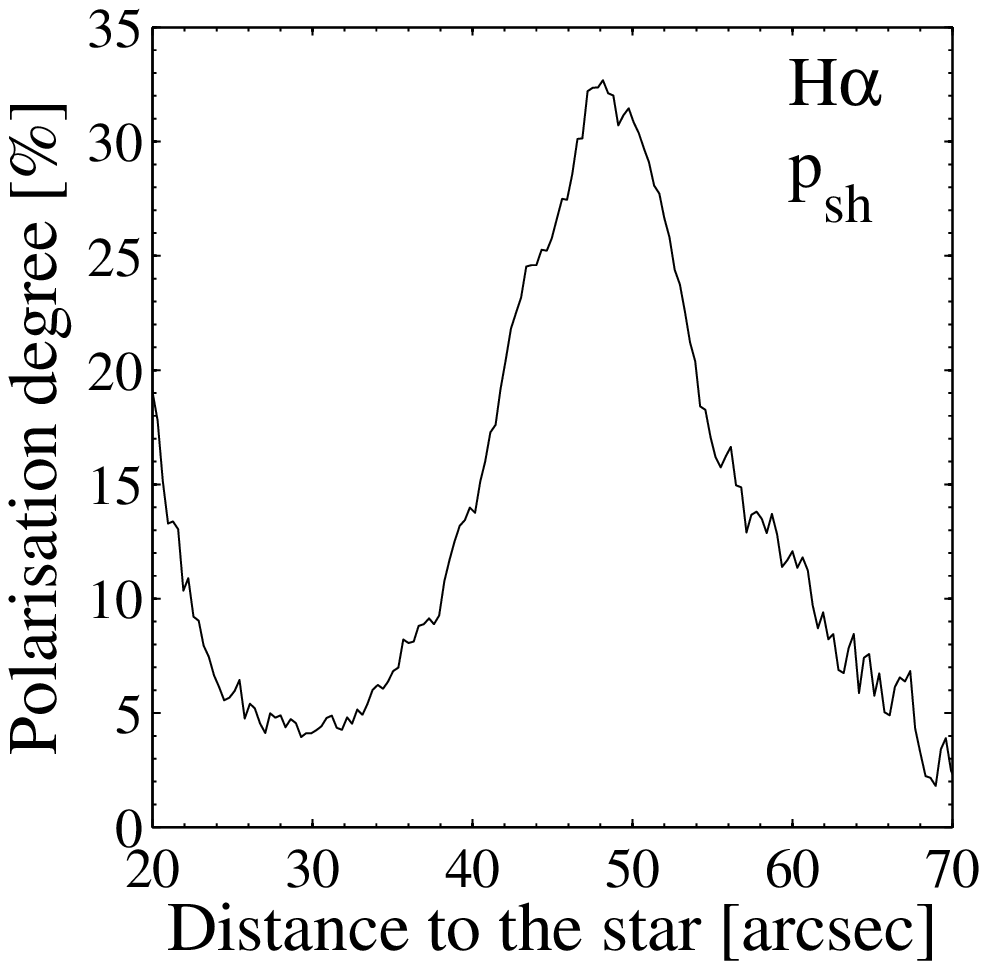}
      \caption{AARPs from the EFOSC2 observations of U~Ant. \emph{Top row, left to right:} the total intensity ($I_{sh}$) in the Str-y, F59, and H$\alpha$ filters, respectively. \emph{Middle row, left to right:} the polarised intensity ($P_{sh}$) in the Str-y, F59, and H$\alpha$ filters, respectively.  \emph{Bottom row, left to right:} The polarisation degree ($p_{sh}$) in the Str-y, F59, and H$\alpha$ filters, respectively. The lines are: the observed data (solid), fits to the individual shells (dotted), and the sum of the individual fits (dashed). See text for details.}
              \label{AARprofs}
    \end{figure*}

Figure~\ref{EFOSC_images} shows the total flux, $I_{sh}$, and polarised flux, $P_{sh}$, images in the Str-y, F59, and H$\alpha$ filters. AARPs of the $I_{sh}$, $P_{sh}$, and the polarisation degree, $p_{sh}$, are shown in Fig.~\ref{AARprofs}. 

The morphology in the H$\alpha$ and Str-y $I_{sh}$ images differs significantly from that in the F59 filter. The former both show a circular disk of $\approx\,$50{\arcsec} radius. The $I_{sh}$ AARPs in both filters show that the intensity remains relatively constant out to a radius of $\approx\,$50{\arcsec} followed by a tail of declining intensity. GD2003 showed that such a brightness distribution is consistent with scattering of stellar light in a detached shell of dust, assuming that the dust has a preferential scattering efficiency in the forward direction. The irregularities in the AARPs are likely due to a few bright arcs present in the images closer to the star. However, a small peak at $\approx\,$43{\arcsec} indicates the presence of an additional shell, and an extended arc at this distance from the star is clearly seen in both filters in the south-east quadrant.

The $I_{\rm{sh}}$ AARP of the F59 image shows a much more pronounced peak at $\approx\,$43\arcsec, while the tail at $\approx\,$48\arcsec is relatively weak. Since the F59 filter may contain a contribution from stellar light scattered in the NaD resonance lines, while the H$\alpha$ and Str-y filters are dominated by dust scattered light, this confirms the results by GD2003, where shell 3 is dominated by line scattering, while shell 4 is dominated by dust scattering. 

Only light that is scattered at an angle of  $\approx$\,90$^{\circ}$ will be strongly polarised. The polarisation due to the circumstellar medium will hence be strongest in the plane of the sky that goes through the star. Detached, spherical shells appear as ring-like structures in such images, directly revealing the spatial structure of the shells. The images in the polarised flux $P_{sh}$ in all filters clearly show a nearly perfect circular geometry. The corresponding AARPs show the presence of two shells at $\approx\,$43{\arcsec} and $\approx\,$48{\arcsec}, corresponding to the positions of shells 3 and 4 in GD2003, but GD2003 did not detect shell 3 in the polarised light images. These AARPs make it possible to determine the locations and the widths of shells 3 and 4. Fits to the AARPs of the $P_{sh}$ images in all filters show that the total polarised flux is dominated by shell 4 (see Sect.~\ref{shellstruct}). Shell 4 also clearly dominates the AARPs of the $p_{sh}$ images, the degree of polarisation reaching a maximum of 25-30\% in all three filters at a radius of $\approx$50\arcsec, confirming that shell 4 is dominated by stellar light scattered by dust. Since $p_{sh}$ is derived $P_{sh}$/$I_{sh}$, the outer parts (where $I_{sh}$ and $P_{sh}\approx0$) become less reliable. Hence, the picture of a shell of gas at the position of shell 3 containing a small amount of dust, and a detached shell dominated by dust with almost no gas at the position of shell 4 is strengthened. 

Structure inside shell 3 is apparent in the F59 image (but not in the others, including the F77 image of GD2003). The positions of the peaks in the AARP are consistent with shells 1 and 2 in GD2001 at 25{\arcsec} and 37{\arcsec}. It is not clear, however, whether the observed structure is due to clumpy structures in shells 3 and/or 4, or due to additional detached shells closer to the star, see Sect.~\ref{onetwo}. 


\subsection{Scattering in detached gas and dust shells}
\label{dustscat}

In order to understand the origin of detached shells, it is important to accurately determine their physical parameters, such as e.g. shell radius and width, and its density distribution and possible clumpiness. Any determination of physical properties of the shells based on the data presented here requires the proper treatment of the scattering agent properties. For an individual dust grain these are affected by, e.g., the grain composition, size, and shape. The observed brightness distributions (due to scattering by a large number of grains) are affected by the grain size distribution, the density distribution within the shell, and the size and width of the shell. 

The scattering by the grains can be calculated using Mie theory, given the optical constants of the grains. If the grain size is small compared to the scattering wavelength ($a/\lambda \ll 0.16$), the condition for Rayleigh scattering is fulfilled and the scattering can be assumed to be isotropic. However, typical grain sizes are $\approx$\,0.1\,$\mu$m, i.e. comparable to the wavelengths in our data, and asymmetric scattering becomes important.  In order to determine the amount of scattering in a particular direction (i.e. the probability for scattering in a direction $\theta$ with respect to the forward direction), we use an analytical expression (based on light scattered in the interstellar medium; Henyey \& Greenstein~\cite{henyeyco1941})

\begin{equation}
\label{e:Ptheta}
P(\theta)={{1-g^2}\over{(1+g^2-2g\,\rm{cos}\theta)^{3/2}}}.
\end{equation}

\noindent where $g$ is the scattering asymmetry parameter defined as

\begin{equation}
\label{e:g}
g={{\int_0^\pi \! I(\theta)\,\rm{cos}\theta\,\rm{sin}\theta\,d\theta}\over{\int_0^\pi \! I(\theta)\,\rm{sin}\theta\,d\theta}},
\end{equation}

\noindent
where $I(\theta)$ is the scattering phase function. For $g$\,=\,0 the scattering is isotropic, while for $g$\,=\,1 all light is scattered in the forward direction and for $g$\,=\,$-1$ all light is scattered backwards. Draine (\cite{draine2003}) re-examined the expression by Henyey \& Greenstein and concluded that it is valid at wavelengths between $0.4\,\mu$m and $1.0\,\mu$m, i.e., the wavelength range of interest here. The amount of polarisation as a function of the scattering angle is given by

\begin{equation}
\label{e:pol}
p={{1-\rm{cos}^2\theta}\over{1+\rm{cos}^2\theta}}
\end{equation}

\begin{figure*}[t]
   \centering
   \includegraphics[width=6cm]{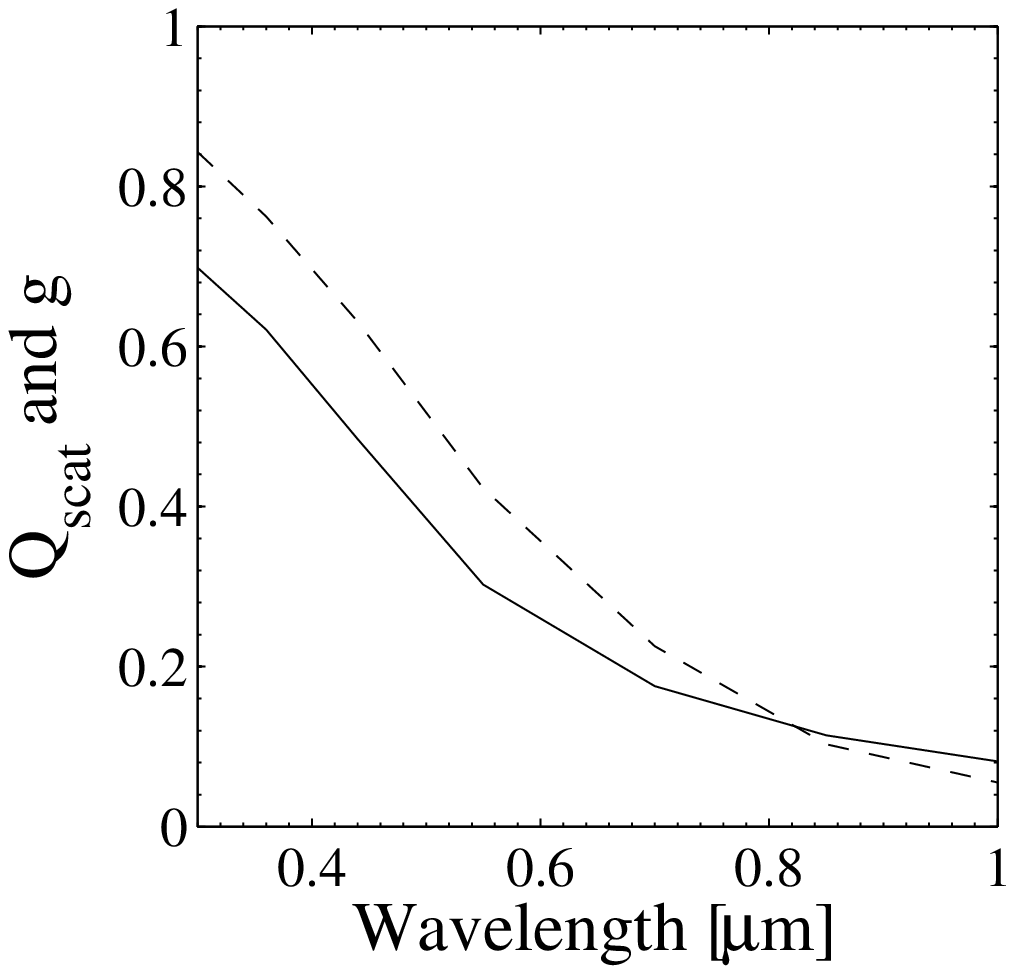}
   \includegraphics[width=6cm]{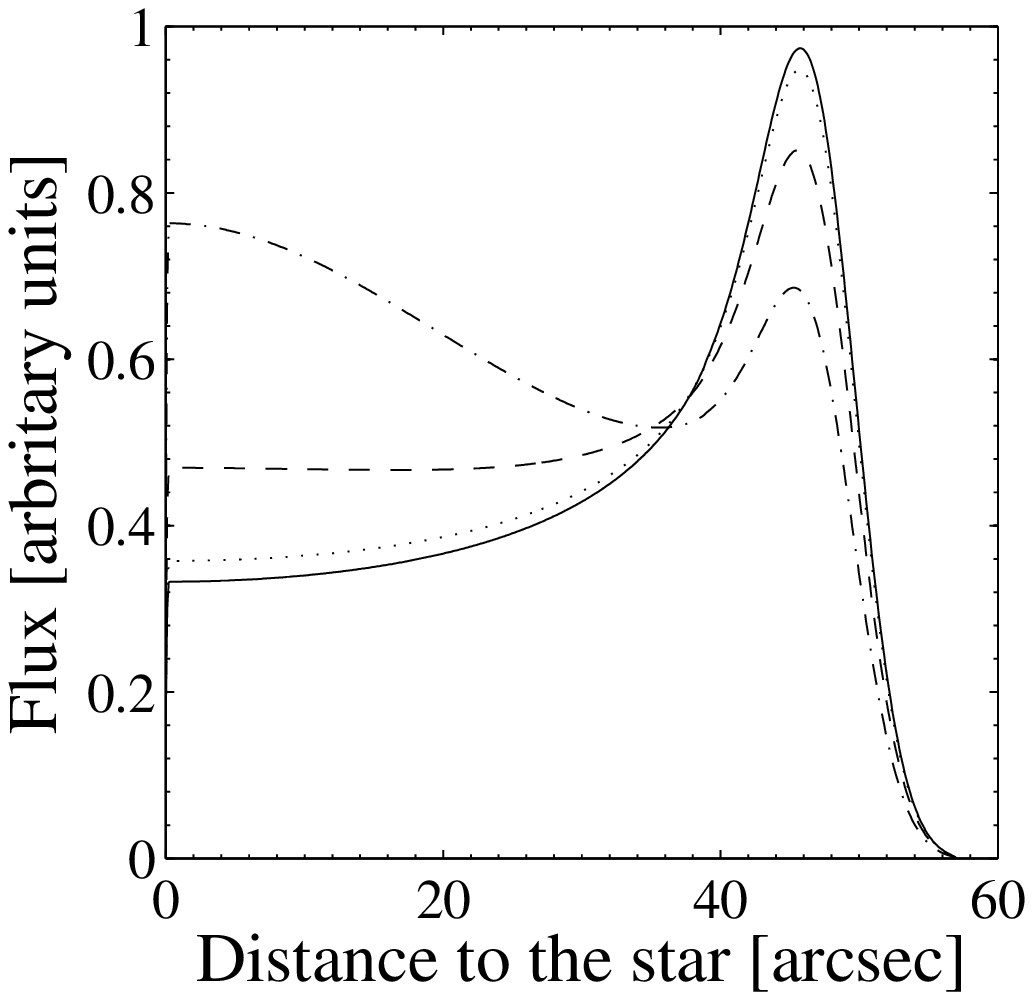}
   \includegraphics[width=6cm]{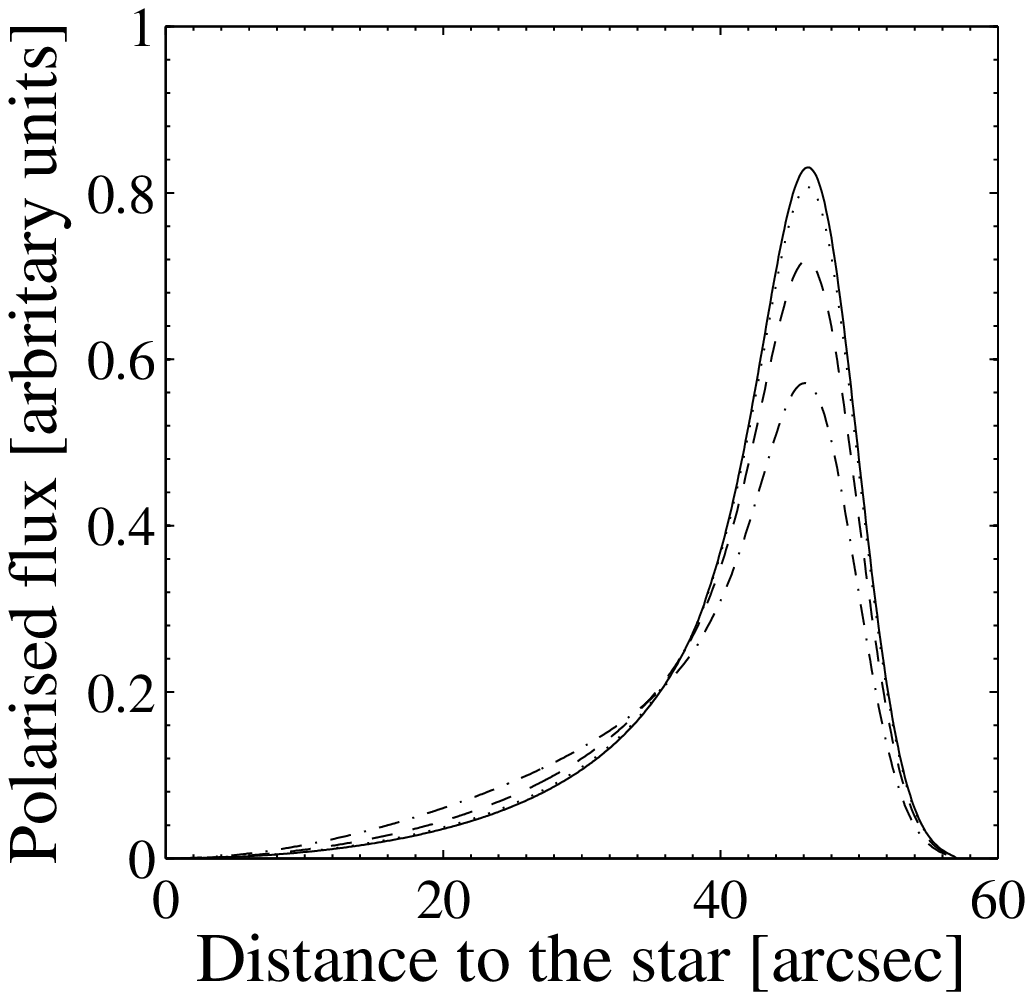}
   \caption{\emph{Left:}The scattering efficiency $Q_{\rm{scat}}$ (dashed line) and the scattering asymmetry parameter $g$ (solid line) vs. wavelength for 0.1\,$\mu$m-sized spherical grains. \emph{Middle:} Model total brightness distributions using the calculated scattering asymmetry parameters for the Str-y filter (dot-dashed line), F59 filter (dashed line), H$\alpha$ filter (dotted line), and for isotropic scattering (solid line). \emph{Right:} The corresponding model brightness distributions for the polarised light using the same units as for the total flux (see text for details).}
   \label{qscatg}
\end{figure*}

The left panel of Fig.~\ref{qscatg} shows the scattering efficiency $Q_{\rm{scat}}$ of dust grains and the scattering parameter $g$ vs. wavelength for amorphous carbon grains (Suh~\cite{suh2000}) assuming spherical grains with radii of 0.1\,$\mu$m. Equation~\ref{e:Ptheta} can then be used to derive radial brightness distributions for different wavelengths assuming single scattering. The middle panel of Fig.~\ref{qscatg} shows the expected brightness distributions for optically thin scattering by dust in a detached shell located at $R$\,=\,48{\arcsec} from the star. They are given for complete isotropic scattering ($g$\,=\,0) and for the calculated $g$ for the respective filters. The density distribution of the detached shell along the radial direction is assumed to follow a Gaussian distribution with a FWHM of $\Delta R=7$\arcsec\, peaking at $R$. The grains have a constant size of $0.1\,\mu$m. The right panel of Fig.~\ref{qscatg} shows the corresponding profiles for the polarised light.

For all cases the limb brightening due to the narrow shell is apparent at $\approx48$\arcsec (hereafter referred to as the `peak'). However, as forward scattering becomes more effective for shorter wavelengths, the brightness profile at line of sights closer to the star increases compared to the peak intensity. Note that the amount of forward scattering does not have an effect on the position of the peak or the sharp decrease at larger radii. However, the density distribution within the shell affects the shape of the peak. A sharp inner edge moves the position of the peak somewhat closer to the star, while a sharp outer edge leads to a sharper decline at large radii. The decline in brightness at large radii follows to good approximation a Gaussian distribution and it gives reasonable estimates of the radius and width of the shell (Mauron \& Huggins 2000). However, fitting a Gaussian distribution to the peak and the tail results in a slightly underestimated radius $R$ (by $\approx$8\%) and an overestimated FWHM of the shell (by $\approx$20\%). Hence, the detailed grain size distribution and the density distribution across the shell are the main uncertainties in determining the shell radius and width from the observed brightness distributions.

In the case of line scattering we use Equ \ref{e:Ptheta} with $g$\,=\,0 for calculating the brightness distribution.


\subsection{Decomposition of the AARPs}
\label{shellstruct}

In order to derive quantitative results for the individual shells it is necessary to decompose the AARPs obtained from the observed images. The images in polarised light best show the spatial structure of the dust, and we use Equ.~\ref{e:Ptheta} and~\ref{e:pol} to derive theoretical AARPs for the polarised scattered light. The scattering asymmetry parameter $g$ is determined for each filter using Mie theory and assuming spherical, 0.1$\,\mu$m sized carbon grains (Suh~\cite{suh2000}). A Gaussian density distribution of the dust as described above is assumed. The model distributions are fit to the peaks in the $P_{\rm{sh}}$ AARPs, hence determining the radius and widths of shells 3 and 4. Equ.~\ref{e:Ptheta} is then used to calculate model profiles of the total intensity $I_{\rm{sh}}$. In the F59 filter shell 3 is assumed to be dominated by scattering in lines, and hence isotropic scattering is assumed in this filter ($g$\,=\,0). The dust contribution to $I_{\rm{sh}}$ from shell 3 in this filter is neglected. The dotted lines in Fig.~\ref{AARprofs} show the individual brightness distributions. Tables~\ref{efoscpolres}~and~\ref{efosctotres} show the results of fitting the brightness distributions to the polarised and total flux AARPs, respectively. For the latter, the shell radii and sizes determined from the polarisation data are used. Hence, we confirm the detection of detached shells at 43\arcsec ( shell 3) and 50\arcsec (shell 4) from the star. The width of shell 3 is $\approx\,$2\farcs2, while shell 4 is a factor of 3 broader. The results for the F77 filter in Table~\ref{efosctotres} are taken from GD2003. The fluxes for the individual shells in Table~\ref{efosctotres} are derived by summing the model brightness distributions over all angles. The total fluxes for all shells are calculated by summing the observed profiles over all angles (labelled `total' in Table~\ref{efosctotres}).

In all measurements probing the contribution to the flux from dust scattered light (i.e. the polarisation measurements and the $I_{\rm{sh}}$ images in the Str-y and H$\alpha$ filters), shell 4 clearly dominates over shell 3. The contribution to the total flux from shell 3 is only comparable to that from shell 4 in the $I_{\rm{sh}}$ images in the F59 filter. This is likely due to the added contribution of line scattered light in shell 3 in this filter.

The theoretical brightness distributions do not perfectly fit the observed AARPs. Artefacts in the images (e.g., diffraction spikes), the uncertainty due to the psf subtraction, and a possible clumpy structure in the detached shells are likely to affect the AARPs, in particular in the inner parts of the images. The assumption of a constant grain-size in the model AARPs and a Gaussian density distribution across the shells also affects the model brightness distributions. The gradual decline at larger radii could also be a kinematic effect, e.g., due to grains of different size travelling at different velocities. The fit to the $I_{sh}$ AARP in the F59 filter is particularly poor. This is partly due to the presence of structures inside shell 3, but there is most likely also an effect of the NaD line scattering being, at least partially, optically thick (see below). Also, the FWHM of shell 3 in the $I_{\rm{sh}}$ AARP seems to be wider than the one determined from the $P_{\rm{sh}}$ AARP. This is likely due to the small amount of dust present in shell 3.

\begin{table*}[th]
\caption{The results of decomposing the $P_{sh}$ AARP data into shell brightness distributions. $R$ and $\Delta R$ are the radius and FWHM of the Gaussian density distribution assumed for the shell, $I\rm{_{peak}}$ and $F\rm{_{tot}}$ are peak intensity and total flux of the brightness distributions fitted to the data. Pol. deg. is the peak polarisation degree at the position of the shells.}
\label{efoscpolres}
\centering
\begin{tabular}{l c c c c c c}
\hline\hline
Filter		& shell	& $R$		& $\Delta R$ 	& $I\rm{_{peak}}$& $F\rm{_{tot}}$ & pol. deg. \\
		&		& [\arcsec]	& [\arcsec]	& [$10^{-16}\,\rm{erg\,s^{-1}\,cm^{-2}\,\arcsec^{-2}}$]& [$10^{-13}\,\rm{erg\,s^{-1}\,cm^{-2}}$] & [\%] \\
\hline
Str y		& 3 & 43.6 & 1.9 & 1.6 & \phantom{1}2.4 & 23\\
		& 4 & 49.6 & 5.4 & 3.7 & 12.0 & 26\\
\hline
F59		& 3 & 43.4 & 2.4 & 0.3 &  \phantom{1}0.5 & \phantom{1}8\\
		& 4 & 49.6 & 7.1 & 0.9 &  \phantom{1}3.4 & 24\\
\hline
H$\alpha$& 3 & 43.4 & 2.4 & 0.5 &  \phantom{1}0.8 & 25\\
		& 4 & 49.8 & 7.1 & 2.0 &  \phantom{1}7.1 & 33 \\
\hline\hline
\end{tabular}
\end{table*}

\begin{table*}[th]
\caption{The results of decomposing the $I_{sh}$ AARP data into shell brightness distributions. $I\rm{_{peak}}$, $F\rm{_{tot}}$, and ${S_{\rm{av}}}$ are the peak intensity, total flux, and average flux density of the shell brightness distribution fits (see text for details). The values for the total of all shells are obtained by integrating the observed brightness distributions. The numbers within brackets in the F59 and F77 filters in the flux column give the percentage of flux that may come from dust, based on a linear inter- and extrapolation between the Str-y and H$\alpha$ filters. ${F_{\rm{sc}}} / {F_{\star}}$ gives the ratio between the circumstellar and stellar fluxes.}
\label{efosctotres}
\centering
\begin{tabular}{l c c c c c}
\hline\hline
Filter		& shell		& $I\rm{_{peak}}$& $F\rm{_{tot}}$ & ${S_{\rm{av}}}$ &${F_{\rm{sc}}} / {F_{\star}}^a$ \\
		&			& [$10^{-16}\,\rm{erg\,s^{-1}\,cm^{-2}\,\arcsec^{-2}}$]& [$10^{-13}\,\rm{erg\,s^{-1}\,cm^{-2}}$] & [$10^{-14}\,\rm{erg\,s^{-1}\,cm^{-2}\,\AA^{-1} }$] &[$10^{-4}$]\\
\hline
Str y 		& 3 	 		& \phantom{1}5.7 		& \phantom{1}16.1		 			& \phantom{1}0.9	& \phantom{1}4.1 	\\
		& 4 	 		& 14.2 				& \phantom{1}80.1					& \phantom{1}4.4	& 20.1 			\\
		& total    		&        				& 144.4 							& \phantom{1}7.9	& 36.3 			\\	
\hline
F59		& 3 	 		& \phantom{1}8.7 		& \phantom{1}20.5 \phantom{1}(19\%)					& \phantom{1}3.9 	& 20.5 			\\
		& 4 	 		& \phantom{1}4.0 		& \phantom{1}24.2 \phantom{1}(83\%) 					& \phantom{1}4.6	& 24.2 			\\
		& total     		&       				& \phantom{1}84.7 \phantom{1}(43\%)	& 16.3			& 84.7 			\\
\hline
H$\alpha$& 3 	 		& \phantom{1}2.4 		& \phantom{1}\phantom{1}6.6 			& \phantom{1}0.5	& \phantom{1}1.6 	\\
		& 4 	 		& \phantom{1}6.1 		& \phantom{1}34.9 					& \phantom{1}2.8	& \phantom{1}8.3 			\\
		& total      		&        				& \phantom{1}68.2 					& \phantom{1}5.5	& 16.3 			\\
\hline
F77		& 3 	 		& 		& \phantom{1}14.0 \phantom{1}\phantom{1}(3\%) & \phantom{1}2.8 	& \phantom{1}3.3      	\\
		& 4 	 		&  		& \phantom{1}\phantom{1}3.3 (100\%)	& \phantom{1}0.7	& \phantom{1}0.8 	\\
		&total		&					&\phantom{1}30.0 \phantom{1}(50\%)	& \phantom{1}6.0	& \phantom{1}7.1	\\
\hline\hline
\end{tabular}
\begin{list}{}{}
\item[$^{\rm{a}}$] The fluxes of the star in the filters are: $4.0\times10^{-9}\,\rm{erg\,s^{-1}\,cm^{-2}}$ (Str y), $1.1\times10^{-9}\,\rm{erg\,s^{-1}\,cm^{-2}}$ (F59), $4.2\times10^{-9}\,\rm{erg\,s^{-1}\,cm^{-2}}$ (H$\alpha$), and $4.2\times10^{-9}\,\rm{erg\,s^{-1}\,cm^{-2}}$ (F77).
\end{list}
\end{table*}


\subsection{Kinematic and spatial information from the EMMI data}
\label{expvel}

\begin{figure}[t]
   \centering
   \includegraphics[width=9.5cm]{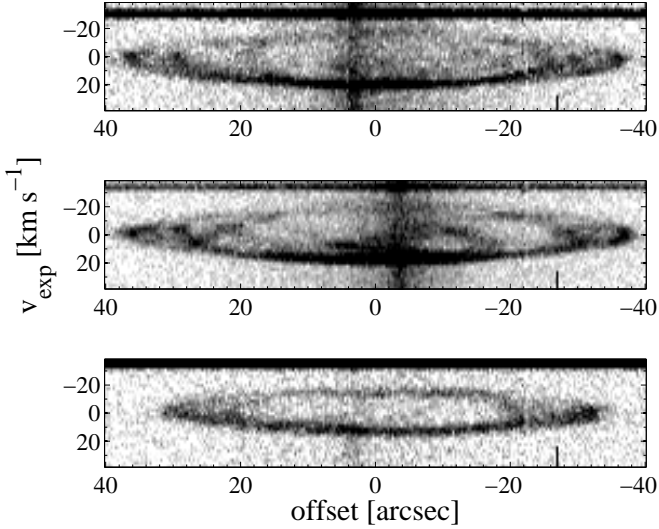}
         \caption{EMMI long-slit spectra towards U~Ant showing the NaD resonance lines. The doublet lines are combined in order to increase the signal-to-noise ratio. The slit is offset from the star by 15\arcsec east (top), 15\arcsec west (middle), and 25\arcsec east (bottom). The velocity scale is given with respect to the systemic velocity.}
              \label{EMMI_Na}
    \end{figure}

\begin{figure}[t]
   \centering
   \includegraphics[width=9.5cm]{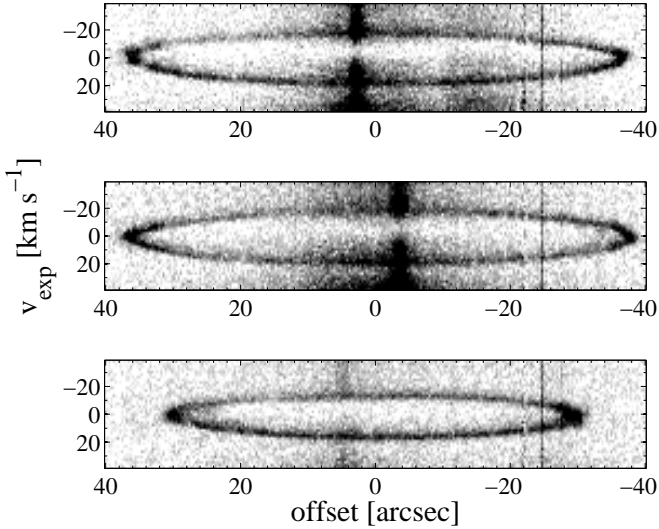}
      \caption{Same as Fig.~\ref{EMMI_Na}~but for the KI resonance line at 769.9 nm.}
              \label{EMMI_K}
    \end{figure}

Examples of the long-slit NaD and KI line data taken with EMMI are shown in Figs~\ref{EMMI_Na} and~\ref{EMMI_K}, respectively. The ellipse-like shape of the lines due to the expansion of a geometrically, essentially spherical thin circumstellar shell can clearly be seen. As mentioned above, the NaD lines show a more complicated morphology, revealing additional structure closer to the star. However, we focus the analysis on the dominating shell. Ellipses were fit to the data (obtained by making cuts along the dispersion axis and fitting Gaussian profiles to the line intensity distribution) using the conical representation of an ellipse

\begin{equation}
\label{coneellipse}
A\cdot x^2+B\cdot xy + C\cdot y^2 + D\cdot x + E\cdot y = 1,
\end{equation}

\noindent
where $x$ and $y$ are the points of the ellipse along the spatial and dispersion axes, respectively. The parameters $A$, $B$, $C$, $D$, and $E$ were determined by fitting the the ellipse to the centres of the Gaussian profiles using a least-squares-fit method. The length $r_{el}$ and width $v_{el}$ of the fitted ellipse are then given by (in arcseconds and $\rm{km\,s^{-1}}$, respectively) 

\begin{equation}
\label{ellipser}
r_{el}=pixscale\times\left({1\over A} + {{D^2}\over{4A^2}} + {{E^2}\over{4AC}}\right)^{{0.5}},
\end{equation}

\begin{equation}
\label{ellipsev}
v_{el}={c\cdot d\over {\lambda}}\times\left({1\over C} + {{D^2}\over{4AC}} + {{E^2}\over{4C^2}}\right)^{{0.5}},
\end{equation}

\noindent
where $c$ is the speed of light, $d$ the dispersion in $\AA\,pix^{-1}$, and $\lambda$ the wavelength of the line. The ellipse parameters are corrected for the projection and hence converted to the size $R$  and expansion velocity $v_{\rm{exp}}$ of the shell. In the case of NaD, the doublet lines were combined in order to increase the signal-to-noise ratio. The F77 filter is wide enough to also detect the line at 766.5~nm, but it is (significantly) weaker than the 769.9 nm line. In addition to the radius of the shell, a cut through the data along the spatial axis at the systemic velocity gives an estimate of the width of the shell $\Delta R$. The resulting parameters from the different ellipse fits and the resulting averages for the shell size, shell width (the shell width determined in the EMMI data is the FWHM of the line along the spatial axis), and expansion velocity are given in Table~\ref{emmires}. 

The EMMI data is dominated by scattering in a shell with a radius of $\approx\,$40{\arcsec} and a width of $\approx\,$3{\arcsec}, that expands at a velocity of 19.5\,$\rm{km\,s^{-1}}$. Although the fit of an ellipse to the data is generally very good, effects due to the physical width of the slit, and the resolved spatial structure of the shell result in deviations from a perfect ellipse at the end points. This causes the measured size of the shell to be systematically smaller (by $\approx$\,2{\arcsec}) than suggested by the peaks measured directly in the data.  Taking this into consideration, the results are fully consistent with the EFOSC2 and CO radio line data on shell 3. The expansion velocity determined from the resonance line data is in excellent agreement with that obtained from models of the CO emission lines, 19.0\,$\rm{km\,s^{-1}}$ (Sch\"oier et al.~\cite{schoieretal2005}).

The brightness of the red-shifted side of the shell (i.e., the rear part as seen from Earth) is clearly brighter than the front in the NaD data. The same effect can be seen to a lesser extent in the KI data. This most likely is an optical depth effect. A higher optical depth in the NaD lines leads to increased reflection of the light on the red-shifted side of the shell, while more of the light is scattered away from the line of sight on the blue-shifted side. 

Finally, the EMMI data taken through the F59 filter also shows structures inside shell 3 in the NaD line, but ellipses are not apparent (see Fig.~\ref{EMMI_Na}). A cut along the spatial axis at the systemic velocity reveals peaks also at 23{\arcsec} and 34{\arcsec} (corrected for projection effects), at the positions of the tentative shells 1 and 2. This structure is not seen in the F77 filter data, i.e., in the KI line.

\begin{table}[t]
\caption{The results of the EMMI data. The NaD doublet lines were added and averaged before fitting an ellipse (see text for details). $\sigma_R$ and $\sigma_{v,\rm{{exp}}}$ are the uncertainties in the radius $R$ and the expansion velocity $v\rm{_{exp}}$ of the shell, repectively. $\Delta R$ is the shell width (measured as the FWHM of the line along the spatial axis). The results are corrected for the projection.}
\label{emmires}
\centering
\begin{tabular}{l l c c c c c }
\hline\hline
\multicolumn{2}{l}{slit	position}& $R$		& $\sigma_R$ 	& $\Delta R$ & $v\rm{_{exp}}$& $\sigma_{v,\rm{{exp}}}$ \\
		& & [\arcsec]	& [\arcsec]	& [\arcsec]	&[$\rm{km\,s^{-1}}$]& [$\rm{km\,s^{-1}}$] \\
\hline
\multicolumn{2}{l}{F59 filter:} & & & & &\\
 15\arcsec~E &(comb)	& 38.5      	& 2.0	 & 3.3 	& 20.2 	& 1.6	 \\
 15\arcsec~W &(comb)	& 40.7 	& 2.0 & 3.3	& 19.5 	& 1.6 \\
 25\arcsec~E &(comb)	& 39.9 	& 3.2 & 3.5	& 17.8 	& 1.6 \\
 all 		&	& 39.7 	& 1.4 & 3.5	& 19.2 	& 0.9\\
\hline
\multicolumn{2}{l}{F77 filter:} & & & & &\\
15\arcsec~E &(769.9) 	& 40.3 	& 2.0 & 1.8	& 19.6 	& 1.8 \\
 15\arcsec~E &(769.9) 	& 40.4 	& 2.0 & 2.2	& 19.6 	& 0.8 \\
 15\arcsec~E &(766.5) 	& 40.2 	& 2.0 & 1.9	& 19.6 	& 1.6 \\
 15\arcsec~W &(769.9) 	& 40.2 	& 2.0 & 2.3	& 20.5 	& 2.0 \\
 15\arcsec~W &(769.9) 	& 41.2 	& 2.0 & 2.7	& 19.6 	& 0.8 \\
 15\arcsec~W &(766.5) 	& 39.3 	& 2.0 & 2.3	& 20.3 	& 1.8 \\
 25\arcsec~E &(769.9) 	& 39.4 	& 2.8 & 3.5	& 19.3 	& 1.8 \\
 25\arcsec~E &(769.9) 	& 39.6 	& 2.8 & 3.7	& 19.4 	& 1.0 \\
 25\arcsec~E &(766.5) 	& 39.6 	& 2.8 & 3.1	& 19.7 	& 1.8 \\
 all 		&	& 40.1 	& 0.8 & 2.6	& 19.7 	& 0.6 \\
\hline\hline
\end{tabular}
\end{table}


\subsection{Line vs. dust scattering}
\label{dustgas}

Previous observations of scattered light in the detached shells around U~Ant (GD2001 and GD2003) were made in filters containing strong resonance lines (except for a tentative detection of shell 4 in a Str\"omgren b filter at 469.0\,nm). Since the widths of the lines are much smaller than the filter width, such observations also contain stellar light scattered by dust, and it is difficult to disentangle the contributions from the different scattering agents. Observations of polarised light mainly measure light scattered by dust (resonance line scattering only accounting for up to $\approx5\%$ of the polarised light; Loskutov \& Ivanov~\cite{loskutovco2007}), and observations with filters that contain no lines make it possible to determine the amount of dust-scattered light, and, together with observations in the ``line''-filters, the amount of line-scattered light.

Table~\ref{efosctotres} gives the flux densities in the different filters. Through linear inter- and extrapolatation of the data in the H$\alpha$ and Str-y filters (assumed to contain contributions only from dust), it is possible to estimate the fractions of flux in the F59 and F77 filters that are due to dust. The result is that $\approx$\,43\% and $\approx$\,50\% of the total fluxes in the F59 and F77 filters, respectively, can be attributed to scattering in dust. The total polarised flux is $\approx$\,7 times higher in shell 4 than in shell 3 in all filters (Table~\ref{efoscpolres}), showing that the dominant contribution to the dust scattered light must come from shell 4. This is in agreement with the results in the polarisation degree, reaching its maximum ($\approx30\%$) at the position of shell 4. Hence, we can conclude that shell~3 is dominated by gas (and line scattering clearly dominates in the F59 ($\approx$\,81\% of the total flux) and F77 ($\approx$\,97\% of the total flux) filter images, while dust dominates in shell 4.


\subsection{The CO shell}
\label{shell3CO}

\begin{figure}[t]
   \centering
   \includegraphics[width=7.5cm,angle=-90]{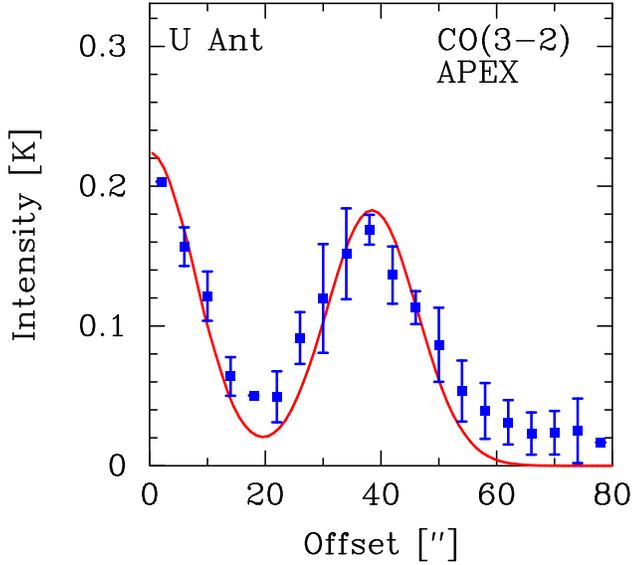}
     \caption{AARP of the APEX CO($J$\,=\,$3-2$) intensity averaged over the central 10\,$\rm{km\,s^{-1}}$ (squares with errorbars). The model CO($J$\,=\,$3-2$) AARP, where the shell position coincides, within the uncertainties, with that of shell 3, is also shown (solid line).}
              \label{apex_rad}
    \end{figure}

Compared to the CO($J$\,=\,$1-0$ and $2-1$) data in Olofsson et al. (\cite{olofssonetal1996}), the APEX data presented here provides a higher angular resolution (18$\arcsec$ beamwidth). Figure~\ref{apex_rad} shows the intensity (averaged over the central 10\,km\,s$^{-1}$) AARP of the new CO($J$\,=\,$3-2$) data. The emissions from the shell and the present-day mass-loss wind are clearly separated. We here model the new CO($J$\,=\,$3-2$) data to show that the CO shell coincides with shell 3. The models are based on the best-fit model presented by Sch\"oier et al. (\cite{schoieretal2005}), slightly adjusted to fit the new APEX data.  The code used in the CO line modelling is described in Sch\"oier \& Olofsson (\cite{schoierco2001}).

Following Sch\"oier et al. (\cite{schoieretal2005}) we adopt a shell width of 1.6$\times$10$^{16}$\,cm (this corresponds to 2\farcs6 at the distance of U~Ant, i.e., close to the shell width as estimated from the line scattering data, which is much smaller than the angular resolution of the APEX CO line data).  The kinetic temperature in the shell is set to 350\,K. The density in the shell is assumed to be constant and defined by the gas shell mass (see below). The expansion velocity of the shell is estimated to be 19.0\,$\rm{km\,s^{-1}}$ by fitting the CO line profiles. For the present-day mass loss we use a mass-loss rate of $1.2\times10^{-8}\,M_{\odot}\,\rm{yr^{-1}}$. This is about a factor of two lower than that obtained by Sch\"oier et al. (\cite{schoieretal2005}). The latter is a result of an average fit to three CO lines, and the SEST CO($J$\,=\,3-2) line is almost a factor of two stronger than the corresponding APEX line. However, it is clear that the APEX line is more reliably calibrated than the SEST line (SEST had a main beam efficiency as low as 25\% at 345\,GHz), and we therefore adjust the mass-loss rate to fit the APEX line. In fact, as can be seen both in Figs~\ref{uant_map} and \ref{apex_rad}, the present-day mass-loss emission is well separated from the detached shell emission, and only contributes to the AARP in the form of a Gaussian profile with the same width as the APEX beam. Therefore, the main goal of the APEX data to verify the position of the detached CO shell is not dependent on the assumption of the present-day mass-loss rate.  The stellar radiation field is a central blackbody with a luminosity of 5800\,$L_{\odot}$ and an effective temperature of 2800\,K. Finally, we adjust the position of the shell to get the best fit to the CO($J$\,=\,$3-2$) intensity AARP. The resulting shell radius is 41$\arcsec$.

The result of the modelling is presented in the form of a model AARP in Fig.~\ref{apex_rad}. The observed AARP is consistent with a model of a detached shell with a radius of 41$\arcsec$, i.e., the CO shell coincides, well within the uncertainties, with shell 3, the shell responsible for the atomic line scattering. The H$_2$ mass of shell 3 is estimated to be $2\times10^{-3}\,M_{\odot}$ assuming a CO abundance with respect to H$_2$ of 10$^{-3}$ (see Table~\ref{shellmass}).

\subsection{The mass of the dust shell}
\label{masses}

All light scattered by dust is likely to be optically thin (see below), and it is therefore possible to do a simple analysis of the observational results, based on an analytical approach described in GD2001. The ratio of the circumstellar flux over the stellar flux is given by

 \begin{equation}
\label{CSSmax}
{{F_{\rm{sc}}} \over {F_{\star}}}= {{\alpha_{\rm{eff}}\Delta \lambda_{\rm{sc}}}\over{\Delta \lambda_{\rm{f}}}}{{N_{\rm{sc}}\sigma_{\rm{sc}}}\over{4\pi R^2_{\rm{sh}}}}<{{\alpha_{\rm{eff}}\Delta \lambda_{\rm{sc}}}\over{{\Delta \lambda_{\rm{f}}}}},
\end{equation}

\noindent 
where $\alpha_{\rm{eff}}$ is ratio of the average stellar flux density in the scattering wavelength range to the average stellar flux density over the filter width, $\Delta \lambda_{\rm{sc}}$ and $\Delta \lambda_{\rm{f}}$ the scattering wavelength range and filter width, respectively, $N_{\rm{sc}}$ the number of scatterers, $\sigma_{\rm{sc}}$ the scattering cross section, and $R_{\rm{sh}}$ the radius of the shell. For dust scattering the maximum ${F_{\rm{sc}}} / {F_{\star}}$ ratio is equal to one ($\alpha_{\rm{eff}}=1$, and $\Delta \lambda_{\rm{sc}}=\Delta \lambda_{\rm{f}}$), and the measured values are all much lower than one (Table~\ref{efosctotres}), hence the scattering is optically thin (assuming a homogeneous medium). 

Assuming single-sized, spherical dust grains the number of scatterers and the scattering cross section are given by

 \begin{equation}
\label{Nscat}
N_{\rm{sc}}={{3M_{\rm{d}}}\over{4\pi \rho_{\rm g}a^3}},
\end{equation}

 \begin{equation}
\label{sigscat}
\sigma_{\rm{sc}}=Q_{\rm{sc}}\pi a^2,
\end{equation}

\noindent
where $M_{\rm{d}}$ is the dust mass in the shell, $\rho_{\rm g}$ the density of a single dust grain, $a$ the radius of a spherical grain, and $Q_{\rm{sc}}$ the grain scattering efficiency. $Q_{\rm{sc}}$ is determined using Mie scattering theory. For simplicity isotropic scattering is assumed. Amorphous carbon dust grains with optical constants from Suh (\cite{suh2000}) were adopted. Typical values for dust in the circumstellar envelopes of AGB carbon stars are $a$\,=\,0.1\,$\mu$m and $\rho_{\rm g}$\,=\,2\,g\,cm$^{-3}$. 

The ${F_{\rm{sc}}} / {F_{\star}}$ ratios are given by the observations. The model brightness distributions do not reproduce the observed $I_{\rm{sh}}$ AARPs well, likely due to the simplified assumption of a constant grain size, the dust density distribution, and a clumpy medium. Hence, using the model brightness distributions to determine ${F_{\rm{sc}}} / {F_{\star}}$ ratios and dust masses individually for shell 3 and 4 would result in very uncertain results. However, shell 4 dominates the contribution from dust (Sects.~\ref{pollight} and~\ref{shellstruct}). We therefore assume that all dust-scattered light comes from shell 4, and the total observed fluxes are used in determining the dust mass (they are corrected for line scattering in the F59 and F77 filters). The resulting dust mass of shell 4 (averaged over all four filters) is 5$\times$10$^{-5}\,M_{\odot}$, Table~\ref{shellmass}. Sch\"oier et al. (\cite{schoieretal2005}) modelled the thermal emission from U~Ant type, and derived a dust mass of $(1.3\,\pm\,1.2)\times10^{-4}\,M_{\odot}$ at a radius of $(5.1\,\pm\,4.0)\times10^{17}\,$cm, consistent with the mass and radius derived for shell 4 here within the considerable uncertainties. Izumiura et al.~(\cite{izumiuraetal1997}) modelled two dust shells observed in high-resolution IRAS images and derived a dust mass for the inner shell of $2.2\times10^{-5}\,M_{\odot}$ at a distance of $2.0\times10^{17}\,$cm, consistent with the results obtained for shell 4 here. GD2003 derived a dust mass for shell 4 of $4\times10^{-6}\,M_{\odot}$, a factor of 10 lower than our estimate, emphasising the uncertainty in estimated dust mass.

\subsection{The mass of the gas shell} 

Using the ${F_{\rm{sc}}} / {F_{\star}}$ ratios for scattering in the NaD and KI lines, corrected for the contribution from dust scattering, it is in principle possible to estimate the gas mass of shell 3 based on the analytical approach from GD2001 (assuming that the contribution to the observed total fluxes from the gas is entirely due to shell 3). The requirement of optically thin scattering sets upper limits also to the ${F_{\rm{sc}}} / {F_{\star}}$ ratios for the KI and NaD lines. For line scattering we assume the scattering wavelength range to be set by the turbulent velocity width. A value typical for circumstellar environments is $1\,\rm{km\,s^{-1}}$. For the strong resonance lines $\alpha_{\rm{eff}}$\,$\approx$\,0.5 and the filter widths of the F59 and F77 filters is $\Delta \lambda_{\rm{f}}$\,=\,5\,nm, putting the upper limit to the ${F_{\rm{sc}}} / {F_{\star}}$ ratio for the KI and NaD  lines at $\approx$\,2.5$\times10^{-4}$ and $\approx$\,4$\times10^{-4}$, respectively (Equ.~\ref{CSSmax}). The observed ${F_{\rm{sc}}} / {F_{\star}}$ ratio is above this limit by a factor of $\approx$\,20 in the F59 filter, while the ${F_{\rm{sc}}} / {F_{\star}}$ ratio in the F77 filter lies close to the limit (only a factor of $\approx$\,3 higher). We conclude that scattering in the NaD lines is optically thick, while scattering in the KI lines may be at least partially optically thin. This is a likely explanation for the different AARPs in the two filters, as well as for the difference in the appearance in the EMMI data. Hence, it is reasonable to expect that we can get at most a lower limit to the shell 3 gas mass using the optically thin approximation for the line-scattered light.

The best gas mass estimate of shell 3 is consequently obtained from the CO line modelling, $2\times10^{-3}\,M_{\odot}$. Assuming that shells 3 and 4 were created at the same time (see below), the derived dust and gas masses give a dust-to-gas mass ratio of $\approx$\,0.02 for the ejected material, not unreasonable for carbon-rich AGB stars.

We can use the optically thin approximation and the observed ${F_{\rm{sc}}} / {F_{\star}}$ ratios (corrected for the dust contribution) to check whether the Na and K results are consistent with the estimated shell gas mass.
Assuming a Boltzmann distribution of the population levels we have for the number of scatterers, 

\begin{equation}
\label{Nlinescat}
N_{\rm{sc}}={{\eta_{\rm{n}}f_{\rm{X}}M_{\rm{sh}}g_{\rm{l}} }\over{\mu\rm{m_H}Z}},
\end{equation}
and for the line scattering cross section,
\begin{equation}
\label{siglinescat}
\sigma_{\rm{sc}}={{1}\over{\Delta\rm{\lambda_{sc}}}}{{1}\over{4\rm{\epsilon_0}}}{{q^2\lambda^2f}\over{m_{\rm{e}}c^2}},
\end{equation}

\noindent
where $\eta_{\rm{n}}$ is the fraction of neutral atoms, $f_{\rm{X}}$ the fractional abundance of species X relative to H, $g_{\rm{l}}$ the statistical weight of the lower level, $E_{\rm{l}}$ the energy of the lower level, $Z$ the partition function, $q$ the charge of an electron, $f$ the oscillator strength of the line, and the other parameters follow usual notations. Assuming solar abundances of Na and K we find that only 13\% and 27\% of the Na and K, respectively,  is required to remain in neutral form to explain the observational results.

\begin{table}
\caption{Derived shell masses based on the EFOSC2 data (dust) and
the APEX CO line data (gas).}
\label{shellmass}
\centering
\begin{tabular}{c c c c}
\hline\hline
shell		& $R$ & $M_{\rm dust}$ & $M_{\rm gas}$ \\
                  & [cm] & [$M_{\odot}$] & [$M_{\odot}$]\\
                  
\hline
3		& 1.7$\times10^{17}$ & & 2$\times10^{-3}$\\
4		& 1.9$\times10^{17}$ & 5$\times10^{-5}$ \\
\hline\hline
\end{tabular}
\end{table}


\section{Discussion and conclusions}
\label{discussion}

\subsection{The identification of shells}
\label{morphs}

Based on the EFOSC2 data we clearly detect two geometrically thin shells around U~Ant with radii of 43\arcsec\,and 50\arcsec\,(shells 3 and 4, respectively). Although shell 4 is relatively weak in the images of total scattered light, it is clearly visible in the images of polarised light. We also confirm the tentative detection of additional structure inside shell 3 in the F59 filter images, a result supported by the EMMI NaD data. We proceed to discuss the tentative shells 1 and 2, and the shells 3 and 4 in separate sections below.

\subsubsection{Shells 1 and 2}
\label{onetwo}

These shells were tentatively introduced by GD2003. The structures attributed to these shells were only seen in the EFOSC2 F59 data. The $I_{sh}$ image shows signs of arcs that may form an essentially circular shell (shell 2). However, this putative shell is not centered on the star. There are also arcs that seem to `connect' this shell and shell 3. The presence of shell 1 was inferred only from the $I_{sh}$ AARP. However, here we provide evidence for structures in the radial ranges of shells 1 and 2 also in the EMMI NaD line data, but these structures do not form complete (or even partial) ellipses as expected from spherical shells centered on the star. Curiously, these structures are not seen in any of the other data. The most reasonable explanation for this is that the optical depth of the NaD line scattering is high (above one) and that of the other scattering agents is low (below one). This is consistent with the asymmetry in the brightness between the red- and blue-shifted sides in the EMMI data. The effect is stronger in the NaD lines, indicating a higher optical depth than in the KI line. The spatial resolution and the dynamic range of the CO data are not enough to identify structures in this radial range. We conclude that there is no strong evidence for the existence of full shells inside shell 3, but structures do exist and their interpretation is not clear. Whether these structures indeed are closer to the star, or whether they are structures within the observed shells is also not clear.

\subsubsection{Shells 3 and 4}
\label{threefour}

Shells 3 and 4 are clearly present as separate entities in the polarisation data at radii of $\approx$\,43{\arcsec} and 50{\arcsec}, respectively. In the $P_{sh}$ images shells 3 and 4 have a large degree of circular symmetry, with slight variations in the intensity indicating a somewhat clumpy structure. There is a gap in shells 3 and 4 in the north-eastern corner of the polarised images, however, this is likely due to artefacts in the images, rather than a gap in the density distribution in the shells. The single-dish CO radio observations do not resolve individual shells, nor do they indicate a multiple-shell structure. However, there are several indications that the CO line emission comes predominantly from shell 3 (Sect.~\ref{shell3CO}). In addition, there is a `hole' in the south-western quadrant in the CO maps (Olofsson et al.~\cite{olofssonetal1996}) and this is seen also in the F59 and F77 images. Based on the flux densities in the different filters and the polarisation data, it is clear that shell 4 essentially consists of only dust, while shell 3 is dominated by gas. 
 
The results of the brightness distribution fits give FWHM widths (of the density distributions of $\approx$\,2{\arcsec} and $\approx$\,7{\arcsec} for shell 3 and 4, respectively. The narrow width of shell 3 is confirmed by the EMMI data, and consistent with the CO line data. Thus, we conclude that the gas shell is geometrically thin ($\Delta R/R$\,$\le$\,0.1), while the external dust shell is wider.

\subsection{The evolution of the detached shells}
\label{evolve}

Assuming a distance of 260 pc to U~Ant (the Hipparcos distance), we derive mean shell sizes of $1.7\times10^{17}$\,cm and $1.9\times10^{17}$\,cm for shells 3 and 4, respectively. The EMMI data shows that shell 3 has an expansion velocity of $\approx$\,19.5\,$\rm{km\,s^{-1}}$, a value consistent with the CO line results. This corresponds to a dynamical shell 3 age of $\approx$\,2700 years, assuming that the shell expansion velocity has been constant with time.

We believe shell 4 to consist of only dust, while shell 3 is dominated by gas. A possible interpretation of this fact is that both shells originate in the same event, but that the dust, moving at a slightly higher velocity, has to a large extent separated from the gas and formed a shell of its own further out. This was already suggested by GD2003. Assuming that both shells were created 2700 years ago, this implies a drift velocity of 2.8$\rm{\,km\,s^{-1}}$ between the gas and the dust, certainly reasonable for winds of AGB stars with high mass-loss rates (Ramstedt et al.~\cite{ramstedtetal2008}). Further, based on the results for the gas mass of shell 3 from the CO line models, $2\times10^{-3}\,M_{\odot}$, and assuming that all the dust in shells 3 and 4 comes from the same event, the dust-to-gas mass ratio of the ejected material is $\approx\,$0.02.

The derivation of the drift velocity assumes that the dust, once separated from the gas shell, moves at a constant velocity. However, when the dust leaves the gas, the drag force due to collisions with the gas disappears, and the radiation pressure from the star again accelerates the grains. The equation of motion for grains that have separated from the gas is

\begin{equation}
\label{radpress}
{{d^2r}\over{dt^2}}={{3LQ_{\rm sc}}\over{16\pi a\rho_{\rm g} c r^2}},
\end{equation}

\noindent
where $r$ is the distance from the star, $L$ the stellar luminosity, $Q_{\rm sc}$, $a$, and $\rho_{\rm g}$ the scattering efficiency, the radius, and the density of a grain, respectively. We solve Equ.~\ref{radpress} numerically, assuming the same grain properties as in Sect.~\ref{masses}, to determine at what time shell 4 must have separated from shell 3 in order for both shells to arrive at their observed distances from the star. For the evolution of the shells after separation, we assume that shell 3 continues to expand with a constant velocity of $\approx$\,19.5\,$\rm{km\,s^{-1}}$, while shell 4 has a velocity of 22.3\,$\rm{km\,s^{-1}}$ at the moment of separation. The solution shows that shell 4 and 3 must have separated only $\approx$\,110 years ago for both shells to reach their current positions. However, the probability of observing the detached dust and gas shells immediately after separation is unlikely, and it is probable that the dust separated at an earlier time. Therefore, we would expect the dust shell to be at a significantly larger distance than what is observed. There could be several reasons why the distance between shell 3 and 4 is smaller than expected. A possible explanation is friction outside the gas shell, due to an earlier stellar wind. Simple estimates suggest, however, that this is not efficient enough. Another possibility is that the grain radius $a$ is greater than the adopted value of 0.1\,$\mu$m. Since the acceleration scales as $a^{-1}$, the predicted distance between the shells would shrink. The large width of the dust shell is possibly attributable to dust grains of different sizes moving with different velocities as well. Finally, our assumption concerning the shells, having constant mass, constant thickness, and a smooth medium, clearly is an oversimplification, as shown by the hydrodynamic simulations by Steffen et al. (\cite{steffenetal1998}), Steffen \& Sch{\"o}nberner (\cite{steffenco2000}), and Mattsson et al. (\cite{mattssonetal2007}). The gradual sweeping up of earlier stellar winds may increase the mass density in the gas shell and thus the friction with time (Sch\"oier et al.~\cite{schoieretal2005}), which could lead to a slower passage of the dust grains through the shell, hence delaying the separation of dust and gas. A realistic modelling of the dust drift in such a system is beyond the scope of this paper.

\subsection{The origin of the detached shells}
\label{origin}

A link between geometrically thin, detached shells and thermal pulses was proposed already 20 years ago, although the exact mechanism remained unclear (Olofsson et al.~\cite{olofssonetal1990}). The shells were suggested to have formed during a period of increased mass-loss rate due to changes in the stellar temperature and luminosity during the thermal pulse. However, hydrodynamical models showed that a mass injection due to a period of high mass loss alone proved not sufficient to form the observed gas shells (Steffen et al.~\cite{steffenetal1998}), as they tended to diffuse. Instead, the interaction of the thermal pulse wind with a previous wind of lower velocity proved to be a more promising mechanism for producing the shells  (Steffen \& Sch\"onberger~\cite{steffenco2000}). Although Steffen \& Sch\"onberger managed to create detached shells in connection with thermal pulses in their models, they relied on empirical mass-loss descriptions, even during the thermal pulse. Mattson et al. (\cite{mattssonetal2007}) for the first time self-consistently modelled the change in stellar parameters, the dynamics in the wind acceleration zone, and the evolution of the expanding circumstellar envelope during a thermal pulse. Their models form detached shells due to an increase in expansion velocity and  mass loss during the pulse. Critical for the formation of detached shells in their models is the effect of the stellar pulsation  characteristics (described as a piston at $\approx$\,0.9 stellar radii), and the amplitude of the change in mass-loss rate and expansion velocity. They find shell densities of the order of $10^{-21}\,\rm{g\,cm^{-3}}$ at distances of 1\,$-$\,2$\times10^{17}$\,cm, and expansion velocities of 18\,$\rm{km\,s^{-1}}$, in good agreement with the values derived from our observations. The resulting relative shell thickness in the model ($\Delta R/R\approx0.01$) is approximately a factor of 10 smaller than what observations show. However, the thickness of a shell is sensitive to model assumptions, such as heating and cooling, isothermal or adiabatic shocks, and the jump in expansion velocity. Observational support for a two-wind interaction scenario was given by radiative transfer modelling of dust and CO radio line emission for seven stars with geometrically thin shells by Sch\"oier et al. (\cite{schoieretal2005}). They found evidence of increasing shell mass and decreasing shell expansion velocities with increasing shell radius. This is consistent with a scenario where an expanding shell collides with a previous, slower wind, sweeping up material and hence slowing down. 

Although a separation of the dust and gas due to different expansion velocities seems a reasonable explanation, a different behaviour of the dust and gas components may also be explained by different creation mechanisms. Steffen et al. (\cite{steffenetal1998}) describe the creation of dust shells as a natural consequence of the large drop in mass-loss rate after the thermal pulse. The decrease in mass-loss rate leads to a sharp decrease of dust condensation during the mass-loss rate minimum after the thermal pulse, in turn leading to an uncoupling of the dust and gas. This increases the dust outflow velocity resulting in the formation of a detached dust shell, i.e. interaction with a previous wind is not necessary to create detached shells of dust.

We conclude that the results from the observations are consistent with a thermal pulse scenario (as described above) taking place about 2700 years ago. A gas/dust shell was created that expanded with a velocity higher than during normal mass-loss conditions. The dust moved with a slightly higher velocity and it remained unaffected by any dust in a previous wind. The gas, on the other hand, interacted with the previous lower velocity wind creating a narrow shell with gradually increasing gas mass. The question arises whether the gas or the dust density distribution follows most closely the mass-loss rate modulation during the ejection. The gas shell width corresponds to a time scale of about 150 years, while the dust shell width corresponds to about 400 years. The gas shell is affected by interaction while the dust shell can be affected by grain-size-dependent expansion velocities. To answer this question requires models that follow in detail the evolution of the gaseous and dusty media. 

We confirm that there is circumstellar structure in the region inside shell 3 (i.e., at projected distances closer to the star), but there is no good evidence that these form shells, i.e., that they can be attributed to additional variations in the mass-loss rate. Higher quality data is required to solve this issue, which remains a point of some concern for the overall interpretation.
Assuming the same expansion velocity for the putative shell 2 as for shell 3 implies an age difference of only 600 years - not in any obvious way consolable with the time scale between consecutive thermal pulses. Multiple shells (or rather arcs) have been observed around e.g. the carbon star IRC+10216 (Mauron \& Huggins~\cite{mauronco1999}; Mauron \& Huggins~\cite{mauronco2000}; Fong et al.~\cite{fongetal2003}), separated by time scales of a few hundred to thousand years. The creation of this type of shells with time scales of less than the interpulse time scale may be explained by density and velocity modulations during a single thermal pulse, in addition to being the consequence of successive pulses (Villaver et al.~\cite{villaveretal2002}).



\begin{acknowledgements}
 The authors acknowledge the financial support from the Swedish Research Council and the Swedish National Space Board. We are deeply indebted to our former colleague Hugo Schwarz who participated in some of the observations presented here. His untimely death in 2006 put an end to a fruitful and enjoyable collaboration. We are also indebted to David Gonz\'alez Delgado for performing the EFOSC2 observations.
\end{acknowledgements}
    


\begin{thebibliography}{}

\bibitem[1979]{bernes1979} Bernes, C.
	1979, A\&A, 73, 67

\bibitem[1995]{blocker1995} Bl\"ocker, T.
	1995, A\&A, 297, 727

\bibitem[1988]{boothroydco1988a} Boothroyd, A.I., \& Sackmann, I.-J.
	1988, ApJ, 328, 632

\bibitem[2001]{delgadoetal2001} Gonz\'alez Delgado, D., Olofsson, H., Schwarz, H.E., Eriksson, K., \& Gustafsson, B.
	2001, A\&A, 372, 885 (GD2001)
	
\bibitem[2003]{delgadoetal2003} Gonz\'alez Delgado, D., Olofsson, H., Schwarz, H.E., Eriksson, et al.
	2003, A\&A, 399, 1021 (GD2003)

\bibitem[2003]{draine2003} Draine, B.T.
	2003, ApJ, 598, 1017 

\bibitem[2003]{fongetal2003}ÊFong, D., Meixner, M., \& Shah, R.Y.
	2003, ApJ, 582, L39

\bibitem[1997]{forestinico1997} Forestini, M., \& Charbonnel, C.
	1997, A\&A Supl. Ser., 123, 241

\bibitem[2003]{gerardco2003} G\'erard, E., \& Le Bertre, T.
	2003, A\&A, 397, L17

\bibitem[1998]{groenewegenetal1998} Groenewegen, M.A.T., Whitelock, P.A., Smith, C.H., \& Kerschbaum, F.
	1998, MNRAS, 291, 18

\bibitem[1996]{habing1996} Habing, H.
	1996, A\&AR, 7, 97

\bibitem[1998]{hashimotoetal1998} Hashimoto, O., Izumiura, H., Kester, D.J.M., \& Bontekoe, Tj.R.
	1998, A\&A, 329, 213

\bibitem[1941]{henyeyco1941} Henyey, L.G., \& Greenstein, J.L.
	1941, ApJ, 93, 70

\bibitem[1996]{izumiuraetal1996} Izumiura, H., Hashimoto, O., Kawara, K., Yamamura, I., \& Waters, L.B.F.M.
	1996, A\&A, 315, L221

\bibitem[1997]{izumiuraetal1997} Izumiura, H., Waters, L.B.F.M., de Jong, T., et al.
	1997, A\&A, 323, 449

\bibitem[2003]{knappetal2003} Knapp, G.R., Pourbaix, D., Platais, I., \& Jorissen, A.
	2003, A\&A, 403, 993

\bibitem[2007]{libertetal2007} Libert, Y., G\'erard, e., \& Le Bertre, T.
	2007, MNRAS, 380, 1161

\bibitem[1999]{lindqvistetal1999} Lindqvist, M., Olofsson, H., Lucas, R., et al.
	1999, A\&A, 351, L1

\bibitem[2007]{loskutovco2007} Loskutov, V.M., \& Ivanov, V.V.
	2007, Ap, 50, 2

\bibitem[2007]{mattssonetal2007} Mattsson, L., H\"ofner, S., \& Herwig, F.
	2007, A\&A, 470, 339

\bibitem[1999]{mauronco1999} Mauron, N., \& Huggins, P.J.
	1999, A\&A, 349, 203
	
\bibitem[2000]{mauronco2000} Mauron, N., \& Huggins, P.J.
	2000, A\&A, 359, 707

\bibitem[1969]{moffat1969} Moffat, A.F.J.
	1969, A\&A, 3, 455

\bibitem[1988]{olofssonetal1988} Olofsson, H., Eriksson, K., \& Gustafsson, B.
	1988, A\&A, 196, L1

\bibitem[1990]{olofssonetal1990} Olofsson, H., Carlstr\"om, U., Eriksson, K., Gustafsson, B., \& Willson, L.-A.
	1990, A\&A, 230, L13

\bibitem[1993]{olofssonetal1993} Olofsson, H., Eriksson, K., Gustafsson, B., \& Carlstr\"om, U.
	1993, ApJS, 87, 267
	
\bibitem[1996]{olofssonetal1996} Olofsson, H., Bergman, P., Eriksson, K., \& Gustafsson, B.
	1996, A\&A, 311, 587

\bibitem[1998]{olofssonetal1998} Olofsson, H., Bergman, P., Lucas, et al.
	1998, A\&A, 330, L1

\bibitem[2000]{olofssonetal2000} Olofsson, H., Bergman, P., Lucas, R., et al.
	2000, A\&A, 353, 583

\bibitem[2001]{polsetal2001} Pols, O.R., Tout, C.A., Lattanzio, J.C., \& Karakas, A.I.
	2001, in Astronomical Society of the Pacific Conference Series, Vol. 229, Evolution of Binary and Multiple Star Systems, ed. P. Podsiadlowski, S. Rappaport, A.R. King, F.D'Antona, \& L. Burder

\bibitem[2008]{ramstedtetal2008} Ramstedt, S., Sch\"oier, F.L., Olofsson, H., \& Lundgren, A.A.
	2008, A\&A, 487, 645

\bibitem[1991]{raveendran1991} Raveendran, A.V.
	1991, A\&A, 243, 453

\bibitem[2001]{schoierco2001} Sch\"oier, F.L., \& Olofsson, H.
	2001, A\&A, 368, 969

\bibitem[2005]{schoieretal2005} Sch\"oier, F.L., Lindqvsit, M., \& Olofsson, H.
	2005, A\&A, 436, 633

\bibitem[1999]{schroderetal1999} Schr\"oder, K.-P., Winters, J.M., Sedlmayr, E.
	1999, A\&A, 349, 898

\bibitem[2001]{schroderco2001} Schr\"oder, K.-P., \& Sedlmayr, E.
	2001, A\&A, 366, 913

\bibitem[2000]{specketal2000} Speck, A.K., Meixner, M., \& Knapp, G.R.
	2000, AJ, 545, L145

\bibitem[2000]{steffenco2000} Steffen, M., \& Sch\"onberger, D.
	2000, A\&A, 357, 180 

\bibitem[1998]{steffenetal1998} Steffen, M., Szczerba, M., \& Sch\"onberger, D.
	1998, A\&A, 337, 149

\bibitem[2000]{suh2000} Suh, K.
	2000, MNRAS, 315, 740

\bibitem[2002]{villaveretal2002} Villaver, E., Garc\'ia-Segura, G., \& Manchado, A.
	2002, ApJ, 571, 880

\bibitem[1998]{wagenhuberco1998} Wagenhuber, J., \& Groenewegen, M.A.T.
	1998, A\&A, 340, 183

\bibitem[2006]{wareingetal2006} Wareing, C.J., Ziljstra, A.A., Speck, A.K., et al.
	2006, MNRAS, 372, L63

\bibitem[1994]{watersetal1994} Waters, L.B.F.M., Loup, C., Kester, D.J.M., Bontekoe, Tj.R., \& de Jong, T.
	1994, A\&A, 281, L1

%
\end{thebibliography}
\end{document}